\documentclass{jfm}
\usepackage{graphicx}
\usepackage{hyperref}
\usepackage{epstopdf, epsfig}
\usepackage{natbib}
\usepackage{tabularx}
\usepackage{subcaption}
\usepackage{amsmath}
\usepackage{bbm}
\usepackage[inline,shortlabels]{enumitem}
\usepackage[normalem]{ulem}
\usepackage[english]{babel}
\usepackage{tikz}
\usepackage{xcolor}
\usepackage{lmodern,pdftexcmds,amsmath}
\colorlet{revision}{black}
\colorlet{revision2}{black}
\colorlet{revision3}{black}

\definecolor{gpltRed}   {HTML}{DC143C}
\definecolor{gpltBlue}  {HTML}{4169E1}
\definecolor{gpltGreen} {HTML}{6B8E23}
\definecolor{gpltGold}  {HTML}{DAA520}
\definecolor{gpltPurple}{HTML}{9932CC}

\definecolor{gpltphase0} {HTML}{AEE56C}
\definecolor{gpltphase30} {HTML}{3D7F6A}
\definecolor{gpltphase60} {HTML}{0F4CBD}
\definecolor{gpltphase90} {HTML}{652499}
\definecolor{gpltphase120} {HTML}{C71585}
\definecolor{gpltphase150} {HTML}{FF0000}

\DeclareMathAlphabet{\mathsfbi}{OT1}{\sfdefault}{bx}{sl}
\DeclareMathVersion{sfletters}
\SetSymbolFont{letters}{sfletters}{OML}{ntxsfmi}{b}{it}

\makeatletter
\newcommand{\mathbfsbilow}[1]{%
  \text{\mathversion{sfletters}$\m@th#1$}%
}

\DeclareRobustCommand{\tensor}[1]{%
  \begingroup
  \ifcat\noexpand #1\relax
    \edef\greek@test{\detokenize{#1}}%
    \edef\greek@test{\expandafter\@cdr\greek@test\@nil}%
    \edef\greek@test{\expandafter\@car\greek@test\@nil}%
    \edef\x{\the\lccode\expandafter`\greek@test}%
    \edef\y{\number\expandafter`\greek@test}%
    \ifnum\x=\y\relax
      \mathbfsbilow{#1}%
    \else
      \mathsfbi{#1}%
    \fi
  \else
    \mathsfbi{#1}%
  \fi
  \endgroup
}
\makeatother

\title{\textcolor{revision3}{Dynamics of an oscillatory boundary layer over a sediment bed in Euler-Lagrange simulations}}
\author{
    Jonathan S. Van Doren\aff{1}
    \and
    M. Houssem Kasbaoui\aff{1}\corresp{\email{houssem.kasbaoui@asu.edu}}}
\shortauthor{J. S. Van Doren, and M. H. Kasbaoui}

\affiliation{
  \aff{1} School for Engineering of Matter, Transport and Energy, Arizona State University, Tempe, AZ 85281, USA.
}

\DeclareRobustCommand\LineStyleSolid{\tikz[baseline]     \draw[]          (0,0.5ex) --++ (12pt,0);}

\begin{document}
\maketitle
\begin{abstract}
    \color{revision}
We investigate the dynamics of an oscillatory boundary layer developing over a bed of collisional and freely evolving sediment grains. We perform Euler-Lagrange simulations at Reynolds numbers $\mathrm{Re}_\delta= 200$, 400, and 800, density ratio $\rho_p/\rho_f = 2.65$, Galileo number $\mathrm{Ga} = 51.9$, maximum Shields numbers from \textcolor{revision3}{$5.60 \times 10^{-2}$ to $2.43 \times 10^{-1}$, based on smooth wall configuration,} and Keulegan-Carpenter number from $134.5$ to $538.0$. We show that the dynamics of the oscillatory boundary layer and sediment bed are strongly coupled due to two mechanisms: (I) bed permeability, which leads to flow penetration deep inside the sediment layer, a slip velocity at the bed-fluid interface, and the expansion of the boundary layer, and (II) particle motion, which leads to rolling-grain ripples at $\mathrm{Re}_\delta = 400$ and $\mathrm{Re}_\delta = 800$. While at $\mathrm{Re}_\delta = 200$ the sediment bed remains static during the entire cycle, the permeability of the bed-fluid interface causes a thickening of the boundary layer. With increasing $\mathrm{Re}_\delta$, the particles become mobile, which leads to rolling-grain ripples at $\mathrm{Re}_\delta = 400$ and suspended sediment at $\mathrm{Re}_\delta = 800$.  Due to their feedback force on the fluid, the mobile sediment particles cause greater velocity fluctuations in the fluid. \textcolor{revision3}{Flow penetration causes a progressive alteration of the fluid velocity gradient near the bed interface, which reduces the Shields number based upon bed shear stress.}
\end{abstract}

{\bf MSC Codes }  {\it(Optional)} Please enter your MSC Codes here

\section{Introduction}

In shallow areas of the ocean, the seafloor \textcolor{revision}{may be} subject to large oscillating pressure gradients and strong shear forces. This causes sediment to become suspended and transported to new locations, where they are deposited as the shear force oscillates. A model flow often used to investigate this process is the oscillatory boundary layer (OBL) problem. \citet{stokesEffectsInternalFriction1855} was amongst the first to address this problem, specifically  in the limit where viscous effects dominate and where the bottom surface is represented as a smooth flat wall. Under these assumptions, \citet{stokesEffectsInternalFriction1855} derived analytical solutions which show the establishment of a boundary layer with characteristic thickness $\delta =\sqrt{2\nu/\omega}$, where $\nu$ is the fluid kinematic viscosity and $\omega$ the angular frequency of the oscillations. Due to the assumption of dominating viscous effects, these solutions apply only in the limit of very small Reynolds numbers $Re_\delta=U_0 \delta/\nu$, where $U_0$ is the velocity amplitude of the oscillations. Later, many researchers investigated the dynamics of oscillatory boundary layers over smooth and rough walls at larger Reynolds numbers, including when the Reynolds number is sufficiently high for turbulence to emerge \citep{akhavanInvestigationTransitionTurbulence1991a, carstensenCoherentStructuresWave2010, carstensenNoteTurbulentSpots2012, costamagnaCoherentStructuresOscillatory2003, fytanidisMeanFlowStructure2021, ghodkeDNSStudyParticlebedturbulence2016, ghodkeRoughnessEffectsSecondorder2018, hinoExperimentsTransitionTurbulence1976, mazzuoliTurbulentSpotsOscillatory2019, mazzuoliInterfaceresolvedDirectNumerical2020, mazzuoliFormationSedimentChains2016, ozdemirDirectNumericalSimulations2014, pedocchiTurbulentKineticEnergy2011, salonNumericalInvestigationStokes2007, sarpkayaCoherentStructuresOscillatory1993, vittoriDirectSimulationTransition1998,vittoriSedimentTransportOscillatory2020}. However, it is unclear whether these results are applicable to seafloors. Unlike the previously studied configurations with impermeable and fixed smooth or rough walls, seafloors are made of sediment particles that together form a porous bed. Depending on the details of the flow over it, the bed may be static, with or without bedforms, and may even evolve dynamically as sediment particles saltate or \textcolor{revision}{become} suspended by the flow \citep{finnParticleBasedModelling2016}. \textcolor{revision}{In this manuscript, we investigate how the dynamics of an OBL couple with those of a bottom collisional and freely evolving sediment bed at increasingly large Reynolds numbers.}

There has been significant effort devoted to the characterization of the boundary layer that develops over a smooth or rough wall under the action of an oscillatory forcing.  Depending on the Reynolds number $Re_{\delta}$, different regimes have been identified, as detailed in \citep{vittoriDirectSimulationTransition1998,pedocchiTurbulentKineticEnergy2011,akhavanInvestigationTransitionTurbulence1991,ozdemirDirectNumericalSimulations2014,fytanidisMeanFlowStructure2021}. To summarize, an OBL developing over an impermeable wall may exhibit four different regimes. 
The laminar regime occurs in smooth, rough, and wavy wall OBL at $\Rey_\delta\lesssim 85$. In this regime, the flow is laminar throughout the oscillation cycle \citep{blondeauxTransizioneIncipienteFondo1979,vittoriDirectSimulationTransition1998,akhavanInvestigationTransitionTurbulence1991}, and is well described \textcolor{revision}{by} the analytical solutions of \citet{stokesEffectsInternalFriction1855}. For $85\lesssim \Rey_\delta \lesssim 550$, the flow is in the disturbed laminar regime \citep{hinoExperimentsTransitionTurbulence1976,jensenTurbulentOscillatoryBoundary1989}. The latter is characterized by the appearance of small amplitude perturbations superimposed upon the Stokes flow \citep{vittoriDirectSimulationTransition1998}. \citet{fytanidisMeanFlowStructure2021} found that the Reynolds number thresholds for this regime \textcolor{revision}{depend strongly} on background disturbances. For $550\lesssim\Rey_\delta\lesssim 3460$, the flow enters \textcolor{revision}{the} intermittent turbulent regime and is characterized by sudden turbulence eruption during the decelerating portion of the oscillatory period before relaminarizing again. Lastly, the turbulent regime occurs for $\Rey_\delta \gtrsim 3460$. In this regime, \citet{jensenTurbulentOscillatoryBoundary1989} show that the OBL presents sustained velocity fluctuations and a logarithmic layer for at least 90 percent of the cycle.
 
{\color{revision}
Note that, with a bottom rough wall, the Reynolds number thresholds between the different regimes may vary considerably, as the flow characteristics depend on additional roughness parameters, such as the Keulegan-Carpenter number $KC = U_0/(\omega d_p)$, where $d_p$ represents the roughnesses size.
\citet{jensenTurbulentOscillatoryBoundary1989} showed that disturbances caused by fixed sandpaper roughness \textcolor{revision2}{lead to a thicker boundary layer and increased turbulence intensity}. Similarly, \citet{xiongBypassTransitionMechanism2020} found that a flow disturbance created by a wall mounted obstacle leads to earlier transition to turbulence, thereby lowering the threshold $\Rey_\delta$ compared to the one found for a smooth wall OBL. Additional studies of roughness effects can be found in \citep{ghodkeDNSStudyParticlebedturbulence2016,ghodkeRoughnessEffectsSecondorder2018,mazzuoliTurbulentSpotsOscillatory2019}.

Permeability may also have a significant effect on the structure of an OBL developing over a particle bed. \citet{conleyVentilatedOscillatoryBoundary1994} performed experiments with a ventilated OBL, that is, an oscillatory boundary layer with periodic transpiration over a permeable wall. They found that the wall shear stress decreases during suction and increases during injection. For a permeable wall, the no-slip condition may no longer hold, as observed by \citet{liuWaveinducedBoundaryLayer1996} and \citet{breugemInfluenceWallPermeability2006}. By comparing simulations of an OBL with experimental data, \citet{meza-valleFlowOscillatoryBoundary2022} showed that a mixed boundary condition at the surface best captures the flow over a permeable wall. A permeable bed allows flow penetration, which in turn creates a Kelven-Helmholtz instability and an inflection point in the fluid velocity within the boundary layer \citep{sparrowEFFECTBEDPERMEABILITY2012,voermansVariationFlowTurbulence2017}. All these modifications caused by the bed permeability make it difficult to estimate the bed shear stress beforehand \citep{yuanExperimentalStudyTurbulent2014}.

The studies discussed thus far considered fixed porous beds. In the case of mobile beds, particle transport may further alter the bed fluid interface. Particle motion leads to the emergence of new regimes, as discussed by \citet{finnRegimesSedimentturbulenceInteraction2016}, who proposed a regime map for sediment-turbulent interactions. In the no-motion regime, the bed remains stationary and acts as previously described.  In the bedform regime, ripples emerge in the particle bed as particles saltate over the surface. The study of \citet{mazzuoliDirectNumericalSimulations2019}, by means of particle-resolved direct numerical simulations (PR-DNS), illustrates this regime. The authors showed the emergence of rolling-grain ripples in an OBL developing over a sediment bed at $\Rey_\delta \sim 100$, density ratio $\rho_p/\rho_f\sim 2.5$, and Galileo number $\mathrm{Ga}\sim 10$. The latter represents the relative effect of gravitational and viscous forces exerted on sediment grains. An earlier study showed that ripples emerge from the interaction between particle rows and recirculation zones in the OBL \citep{mazzuoliFormationSedimentChains2016}.
In the sheet flow regime, the bed-fluid interface recedes due to the formation of a layer of suspended particles \citep{hsuTwoPhaseSedimentTransport2004,odonoghueFlowTunnelMeasurements2004}. \citet{mazzuoliInterfaceresolvedDirectNumerical2020} showed that the eruption of turbulence plays a role in the sediment resuspension. Further, at relatively low values of Shields number, sediment transport may depend on both Shields number and flow acceleration.

Several computational methods can be leveraged to study the dynamics of an oscillatory boundary layer over a sediment bed. Particle-Resolved Direct Numerical Simulation (PR-DNS), in which the boundary layer around each sediment grain is fully resolved, provides the highest fidelity since it requires little to no-modeling \citep{uhlmannImmersedBoundaryMethod2005,apteNumericalMethodFully2009,kempeImprovedImmersedBoundary2012,breugemSecondorderAccurateImmersed2012,kasbaouiHighfidelityMethodologyParticleresolved2025}. However, this results in very high computational cost, as seen from the PR-DNS of \citet{mazzuoliDirectNumericalSimulations2019}. While using smaller computational domains and shorter integration times may reduce computational cost, it also introduces numerical artifacts. For example, the domain considered in the PR-DNS of \citet{mazzuoliInterfaceresolvedDirectNumerical2020} was too small to allow the natural formation of ripples in the sediment bed. Even with this restrictive domain size, the high cost of the PR-DNS of \citet{mazzuoliInterfaceresolvedDirectNumerical2020} limited the integration time to the first 1 to 4  cycles of the OBL, which may not be enough to reach a statistically \textcolor{revision2}{quasi-periodic} state. In contrast, the Euler-Lagrange method provides a good balance between computational cost and fidelity \citep{capecelatroEulerLagrangeStrategy2013,finnParticleBasedModelling2016}. In this approach, flow features and bed dynamics on scales larger than the particle size are well resolved, while the flow on the scale of the particle is modeled. This makes it possible to simulate the dynamics on length and time scales much larger than those accessible in PR-DNS. Earlier studies showed that this approach can reproduce with high fidelity the dynamics of dense fluidized beds \citep{capecelatroEulerLagrangeStrategy2013} and dense slurries \citep{capecelatroEulerianLagrangianModeling2013,arollaTransportModelingSedimenting2015}, including in sheet flow and bedform regimes. More recently, \citet{raoCoarsegrainedModelingSheared2019} showed that predictions with the Euler-Lagrange method for the evolution of sediment bed under both laminar and turbulent shear flow match very well with PR-DNS \citep{kidanemariamInterfaceresolvedDirectNumerical2014} and experiments \citep{aussillousInvestigationMobileGranular2013}. 

In this paper, we study the interplay between an oscillatory boundary layer and a sediment bed made of collisional and freely evolving particles from the laminar regime to the onset of turbulence using Euler-Lagrange simulations. The structure of the manuscript is as follows. In section \ref{sec:governing_equations}, we provide the governing equations that dictate the evolution of the flow and sediment particles. In section \ref{sec:computational_approach}, we provide details on the computations and parameters used in this study. Next, we analyze statistics of the fluid and solid phases in section \ref{sec:particle_bed}, highlighting how sediment bed dynamics couple with those of the oscillatory boundary layer.  Finally, we give concluding remarks in section \ref{sec:conclusion}.
}
\section{Governing equations}\label{sec:governing_equations}

We use the volume-filtering approach of \citet{andersonFluidMechanicalDescription1967} and Euler-Lagrange methodology of \citet{capecelatroEulerLagrangeStrategy2013}  to describe the dynamics of the sediment-laden flow. The carrier phase is an incompressible fluid with density $\rho_f$ and viscosity $\mu_{f}$. The volume-filtered Navier-Stokes equations read
\begin{eqnarray}
    \frac{\partial}{\partial t} (\alpha_{f}  \rho_{f})+\nabla \cdot (\alpha_{f} \rho_{f} \boldsymbol{u}_{f}) &=& 0 \label{eq:continuity},\\
    \frac{\partial}{\partial t}(\alpha_f {\rho_f} \boldsymbol{u}_{f})+\nabla \cdot (\alpha_f {\rho_f} \boldsymbol{u}_{f} \boldsymbol{u}_{f}) &=& \nabla \cdot \left(\boldsymbol{\tau} + \boldsymbol{R}_{\mu}\right)+\alpha_f {\rho_f} \boldsymbol{g} - \boldsymbol{F}_{p}+ \boldsymbol{A}\label{eq:momentum},
\end{eqnarray}
where $\alpha_f$ is the fluid volume fraction, $\boldsymbol{u}_{f}$ is the volume-filtered fluid velocity, ${\boldsymbol{\tau}}=-p\boldsymbol{I}+\mu [\nabla \boldsymbol{u}_{f}+\nabla \boldsymbol{u}_{f}^{T}-\frac{2}{3}(\nabla \cdot \boldsymbol{u}_{f})\boldsymbol{I}]$ is the resolved fluid stress tensor \citet{capecelatroEulerLagrangeStrategy2013}, \textcolor{revision}{$p$ is pressure, which includes the hydrostatic contribution,} $\boldsymbol{g}$ is the gravitational acceleration, and $\boldsymbol{F}_{p}$ is the momentum exchange between the particles and the fluid. The tensor $\boldsymbol{R}_{\mu}$ represents the so-called residual viscous stress tensor. This term arises from filtering the point-wise stress tensor. It includes sub-filter scale terms which require closure. This term is believed to be responsible for the apparent enhanced viscosity observed in viscous fluids containing suspended solid particles. For this reason, \citet{capecelatroEulerLagrangeStrategy2013} proposed a closure using an effective viscosity, which when combined with the effective viscosity model of \citet{gibilaroApparentViscosityFluidized2007} leads to an expression for the residual viscous stress tensor
\begin{equation}
	\boldsymbol{R}_{\mu} = \mu_{f}(\alpha_{f}^{-2.8}-1)[\nabla \boldsymbol{u}_{f}+\nabla \boldsymbol{u}_{f}^{T}-\frac{2}{3}(\nabla \cdot \boldsymbol{u}_{f})\boldsymbol{I}].
\end{equation}
\textcolor{revision3} {We do not use an eddy viscosity model since none of the cases we discuss here lead to fully developed turbulence.}

In order to study the dynamics of an oscillatory boundary layer, we drive the flow using the last term in equation (\ref{eq:momentum}), which  expresses as 
\begin{equation}
   \boldsymbol{A} = \alpha_f \rho_f U_0 \omega \cos(\omega t)\boldsymbol{e}_x. 
\end{equation}
This represents a harmonic pressure gradient forcing with angular frequency $\omega$ and \textcolor{revision}{velocity} amplitude $U_0$. Here, $x$ is the coordinate in the streamwise direction along the unitary vector $\boldsymbol{e}_x$, $y$ is the coordinate in the wall normal direction, $z$ is the coordinate in the \textcolor{revision}{spanwise} direction.

The particles are described in the Lagrangian frame. Following \citet{maxeyEquationMotionSmall1983a}, the equations of motion of a particle ``$i$'' are given by
\begin{eqnarray}
     \frac{d \boldsymbol{x}^i_p}{d t}(t) &=& \boldsymbol{u}^i_p(t)\label{eq:part_pos},\\
    m_p \frac{d \boldsymbol{u}^i_{p}}{d t}(t) &=& \textcolor{revision}{\boldsymbol{f}^{h,i}_{p} + \boldsymbol{f}^{c,i}_p}, \label{eq:lpt_1}\\
    \textcolor{revision}{I_p \frac{d \boldsymbol{\omega}^i_{p}}{d t}(t)} &\textcolor{revision}{=}& \textcolor{revision}{\boldsymbol{T}_{p}^{c,i}} \label{eq:lpt_2}
\end{eqnarray}
\textcolor{revision}{where $\boldsymbol{x}^{i}_p$, $\boldsymbol{u}^{i}_p$, $\boldsymbol{\omega}^{i}_p$, $m_p$, and $I_p$ are the particle position, velocity, angular velocity, mass, and moment of inertia, respectively.} \textcolor{revision}{$\boldsymbol{f}^{h,i}_p$ represents the hydrodynamic force, which is modeled as \citep{vandorenTurbulenceModulationDense2024}}
\begin{equation}
	\textcolor{revision}{\boldsymbol{f}^{h,i}_{p}(t) = V_{p} \left.\nabla \cdot \boldsymbol{\tau}\right|_{\boldsymbol{x}_p^i} + m_p f_d\frac{\boldsymbol{u}_{f}(\boldsymbol{x}^i_p,t)-\boldsymbol{u}^i_{p}}{\tau_{p}}  + \boldsymbol{f}_{p}^{\mathrm{am},i} + \boldsymbol{f}_{p}^{\mathrm{lift}, i}. \label{eq:lpt_model}}
\end{equation}
where $V_{p} = \pi d^{3}_{p}/6$ is the particle volume. The first term on the right hand side of equation (\ref{eq:lpt_model}) represents the effect of the undisturbed flow field \citep{maxeyEquationMotionSmall1983a}. The second term represents the drag force exerted on a particle. Note that, $\tau_p=\rho_p d_p^2/(18\mu)$ is the particle response time and $f_d$ is a drag correction factor. We use the one by \citet{tennetiDragLawMonodisperse2011}, derived from particle-resolved direct numerical simulations, which accounts for both inertial and high particle volume fraction effects. The last two terms on the right hand side of  (\ref{eq:lpt_model}) represent the added mass and Saffman lift \citep{saffmanLiftSmallSphere1965} forces,
{\color{revision3}
\begin{eqnarray}
	\boldsymbol{f}_{p}^{\mathrm{am},i}&=&\frac{1}{2} \alpha_f \rho_f V_p \left (\frac{D \boldsymbol{u}_f(\boldsymbol{x}^i_p,t)}{Dt} -\frac{d \boldsymbol{u}_p^i}{dt}\right)\\
	\boldsymbol{f}_p^{\mathrm{lift},i} &=& 1.615 J \mu_{f} d_p |\boldsymbol{u}^i_{s}|\sqrt{\frac{d_p^2 |\boldsymbol{\omega}^i| \alpha_f\rho_f }{\mu_f}} \frac{ \boldsymbol{u}^i_{s}\times \boldsymbol{\omega}^i}{|\boldsymbol{u}^i_{s}| |\boldsymbol{\omega}^i|}
\end{eqnarray}}
where $\boldsymbol{\omega}^i = \boldsymbol{\omega}_f(\boldsymbol{x^i_p},t)$ is the fluid vorticity at the particle location, $\boldsymbol{u}^i_{s} = \boldsymbol{u}_f(\boldsymbol{x}^i_p,t) - \boldsymbol{u}^i_p$ is the slip velocity, $J$ is a lift correction, which is equal to one in the model from \citet{saffmanLiftSmallSphere1965}.

 \textcolor{revision}{The term $\boldsymbol{f}_p^{c,i}$} represents the collisional force exerted on the particle due to particle-particle and particle-wall collisions. These collision are modeled using the soft sphere model, as detailed in \citep{capecelatroEulerLagrangeStrategy2013}.
{\color{revision}
 Briefly, the force exerted on particle $a$ due to a collision with particle $b$, denoted $\boldsymbol{f}_{p}^\mathrm{c,b\rightarrow a}$, is decomposed into normal and tangential components. The normal component $\boldsymbol{f}_{p,n}^\mathrm{c,b\rightarrow a}$ is modeled as a linearized spring-dashpot system, i.e,
 \begin{equation}
 	\boldsymbol{f}_{p,n}^\mathrm{c,b\rightarrow a} = \begin{cases}
		-k\delta_{ab} \boldsymbol{n}_{ab}-\eta \boldsymbol{u}_{ab,n} & \text{if } |\boldsymbol{x}_{p}^a-\boldsymbol{x}_{p}^b| < 0.5(d_p^a+d_p^b)+\lambda \\
		0 & \mathrm{else}.
	\end{cases}
 \end{equation}
where $\delta_{ab}= 0.5(d_p^a+d_p^b)-|\boldsymbol{x}_{p}^a-\boldsymbol{x}_{p}^b|$ is the overlap between particles $a$ and $b$, $\boldsymbol{n}_{ab}$ is the unit normal vector between the particles, and $\boldsymbol{u}_{ab,n}$ is the normal relative velocity. The parameters $k$ and the $\eta$ are the spring stiffness and dampening factor, respectively.}

\begin{figure} \centering
    \includegraphics[width=5in]{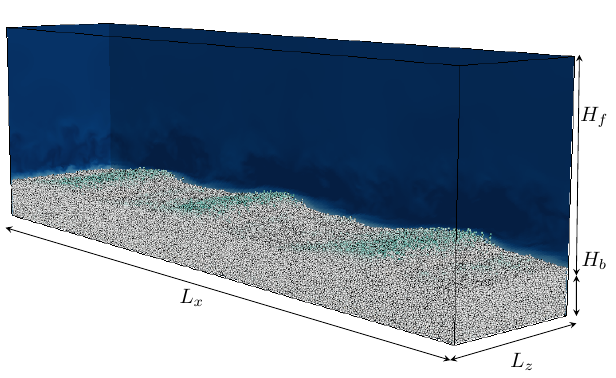}
	\caption{Schematic of the configuration with a bottom sediment bed. The latter is generated in precursor runs where the particles are seeded towards the middle of the domain and allowed to settle on the bottom boundary. }
	\label{fig:configuration}
\end{figure}

Note that the governing equations (\ref{eq:continuity}) and (\ref{eq:momentum}) for the fluid phase are solved in both simulations with particles and without. In the latter case, $\alpha_f=1$ throughout the domain, which recovers the standard Navier-Stokes equations.

\textcolor{revision}{The solid phase dynamics are coupled with the fluid phase dynamics through the momentum exchange field $\boldsymbol{F}_p$, and the volume fraction fields, $\alpha_f$ and $\alpha_p$. We calculate these fields using}
\textcolor{revision}{\begin{eqnarray}
	\boldsymbol{F}_{p}(\boldsymbol{x},t) &=& \sum_{i=1}^N \boldsymbol{f}^{h,i}_{p}(t) g(||\boldsymbol{x}-\boldsymbol{x}^i_{p}||)\label{eq:F_p}\\
	\alpha_{p}(\boldsymbol{x},t) &=& \sum_{i=1}^{N} V_{p} g(||\boldsymbol{x}-\boldsymbol{x}^i_{p}||), \\
	\alpha_{f}(\boldsymbol{x},t) &=& 1 - \alpha_{p}(\boldsymbol{x},t) 
\end{eqnarray}}
In these equations, $g$ represents a Gaussian filter with width \textcolor{revision}{$\delta_{f} = 5 d_{p}$. We study the influence of filter width on fluid statistics in appendix \ref{sec:appendix_a}, and show that the statistics are converged with respect to filter width at $\delta_{f} = 5 d_{p}$.} \textcolor{revision}{Additional details on the computation of these terms and validation of the computational strategy, can be found in \citep{capecelatroEulerLagrangeStrategy2013,capecelatroEulerianLagrangianModeling2013,capecelatroNumericalCharacterizationModeling2014,raoCoarsegrainedModelingSheared2019} and in our recent work  \citep{kasbaouiClusteringEulerEuler2019,kasbaouiTurbulenceModulationSettling2019,shuaiInstabilityDustyVortex2022,shuaiAcceleratedDecayLamb2022,daveMechanismsDragReduction2023,shuaiMergerCorotatingVortices2024,vandorenTurbulenceModulationDense2024}.} \textcolor{revision3}{In appendix \ref{sec:appendix_E}, we further demonstrate the ability of this approach to capture sediment transport accurately. This is done by comparing sediment flow rates obtained with this method in laminar channel flows at varying Shields number with those obtained from PR-DNS \citep{kidanemariamInterfaceresolvedDirectNumerical2014} and laboratory experiments \citep{aussillousInvestigationMobileGranular2013}. Based on the excellent agreement shown in appendix \ref{sec:appendix_E}, we conclude that this Euler-Lagrange method is well suited for this study.
}

\section{Numerical experiments} \label{sec:computational_approach}
\subsection{Configuration}

\begin{table}\color{revision}
    \begin{tabularx}{\linewidth}{XXXXXXXX}
         Cases & $\Rey_{\delta}$ & $\rho_p/\rho_{f}$ & $\mathrm{Ga}$ & $\mathrm{KC}$ & $d_p/\delta$  & $\Theta_\mathrm{max}$\\[1ex]
         1 & 200  & 2.65 & 51.9 & 134.5 & 0.7435 & \textcolor{revision3}{$5.60\,10^{-2}$} \\
         2 & 400  & 2.65 & 51.9 & 269.0 & 0.7435 & \textcolor{revision3}{$1.12\,10^{-1}$} \\
         3 & 800  & 2.65 & 51.9 & 538.0 & 0.7435 & \textcolor{revision3}{$2.43\,10^{-1}$}
    \end{tabularx}
    \caption{Summary of the non-dimensional parameters for the presents runs of an OBL over a sediment bed. \textcolor{revision3}{The maximum Shields number is determined a priori from the smooth wall shear stress as described in \S\ref{sec:fluid_stat}}. }
    \label{tab:Table of parameters}
\end{table}

{\color{revision}
We consider the dynamics of an oscillatory boundary layer over a cohesionless particle bed at three Reynolds number $\Rey_\delta=200$, 400, and 800. A summary of the relevant non-dimensional parameters for each run is listed in table \ref{tab:Table of parameters}. In order to provide a baseline for comparisons with the sediment-laden cases, we also carry out companion simulations at the same Reynolds numbers but with a bottom smooth and impermeable wall instead of a particle bed (see appendix \ref{sec:appendix_SP}). Note that, without the particle bed, the flow at $\Rey_\delta=200$ and 400 is in the laminar regime (see appendix \ref{sec:appendix_SP}). We do not observe a disturbed laminar regime in these simulations due to the wall being smooth and flat and the absence of any external disturbances. \textcolor{revision2}{Without small pertubations the disturbed laminar regime would take much longer to emerge than the simulations times we performed.} The flow at $\Rey_\delta=800$ falls in the intermittent turbulent regime. Comparing the simulations with and without a bottom particle bed helps elucidate the impact of sediment motion, bedforms, and bed permeability on the flow statistics.

In addition to the Reynolds number $\Rey_\delta$, the presence of particles introduces additional dimensionless parameters. These are:
\begin{enumerate*}[(i)]
	\item the density ratio $\rho_p/\rho_f$,
	\item the Galileo number $\mathrm{Ga}=d_{p}\sqrt{(\rho_{p}/\rho_{f}-1)gd_p}/\nu$, and the
	\item and the Keulegan-Carpenter number, $KC=U_{0}/\omega d_{p}$.
\end{enumerate*}
Although not an independent number, the Shields number $\Theta_\mathrm{max}=\tau_{b,\mathrm{max}}/((\rho_{p}-\rho_{f})g d_{p})$, where $\tau_{b,\mathrm{max}}$ is the maximum bed shear stress, is also an important non-dimensional number to consider. The values for each case is shown in table \ref{tab:Table of parameters}. From a dimensional perspective, these cases \textcolor{revision2}{may} correspond to an oscillating flow with period $T=1.75$ s, velocity amplitude varying from $U_0=0.134$ m/s to 0.536 m/s, and sand particles with diameter 550 $\mu$m.

\citet{finnParticleBasedModelling2016} suggest that the regimes of particle transport is determined by the combination of Shields and Galileo numbers. Based on their work and the combination of the present parameters, case 1 ($\Rey_\delta=200$) falls into the ``no motion regime''. Cases 2 ($\Rey_\delta=400$) and 3 ($\Rey_\delta=800$) fall in the gravitational settling regime. We expect particle motion in both of these cases, with notably higher suspended sediment concentration in case 3 compared to case 2. In all these cases, the Keulegan-Carpenter number is very large, which suggests that inertial forces on particles caused by the unsteady flow oscillations are negligible compared to the drag force due to the instantaneous slip.

Figure \ref{fig:configuration} shows a schematic of the computational domain we use for the present simulations. Table \ref{tab:Domain_Parameters} gives a summary of the computational parameters. The domain is long by $250\delta$ in the streamwise direction and by $50\delta$ in the spanswise direction. The domain height in the normal direction is $H_f+H_b$, where $H_f=61.9\delta$ is the height of the clear fluid column and $H_b=16.7\delta $ is the initial bed height. We chose the latter sufficiently deep ($H_b\sim 22d_p$) to accurately account for the flow intrusion within the bed. This gives a total number of particles $N = 6.09\times10^{5}$.

\begin{table}\color{revision2}\centering
	\def\a{0.5in}
	\def\b{1.0in}
    \begin{tabular}{p{\b}p{\a}p{\a}p{\a}p{\a}p{\a}p{\a}p{\a}}
       Cases & $N_x$ & $N_y$ & $N_z$ & $L_x/\delta$ & $L_z/\delta$ & $H_f/\delta$ & $H_b/\delta$\\
       with  & 672 & 211 & 134 & 250 & 50 & 61.9 & 16.7 \\
       without bed    & 64 & 256 & 64 & 50 & 25 & 40 & 0
    \end{tabular}
    \caption{\textcolor{revision}{ Summary of domain parameters.}}
    \label{tab:Domain_Parameters}	
\end{table}

To discretize the governing equations, we use a uniform grid with spacing $\Delta x=d_p/2$ which provides a high resolution of the momentum coupling between and particle Eulerian fields such as particle volume fraction and velocity field. This results in a grid of size $672\times 211\times 134$. \textcolor{revision3}{In section \ref{sec:appendix_d}, we conduct a grid convergence study and show that this discretization is sufficient for $\Rey_\delta = 800$.} The timestep $\Delta t$ is such that $\Delta t/T = 1.79 \times10^{-5}$. This restrictive timestep is imposed by the requirement in the soft-sphere collision model that the bottom layer of the particles must support the weight of the particle bed above them. This is satisfied by ensuring that the spring stiffness in the collision model is sufficiently large to maintain a realistic volume fraction of $0.63$ for a poured-bed \citet{scottDensityRandomClose1969} (more details in \citet{capecelatroEulerLagrangeStrategy2013}). We use periodic boundary conditions in the streamwise and spanwise directions. In the wall-normal direction, we use far-field boundary conditions at the top, and no-slip condition at the bottom. Note that the bottom layer of particles is held fixed, whereas all other particles are free to move according to the evolution equations (\ref{eq:part_pos})--(\ref{eq:lpt_2}). \textcolor{revision3}{Following \citet{charruSedimentTransportBedforms2016}, the particle restitution coefficient is maintained at 0.8 and the particle friction coefficient at 0.4. Note that  \cite{kidanemariamInterfaceresolvedDirectNumerical2014} showed that the precise value of the friction coefficient does not have a significant impact on sediment transport.}

The protocol to initialize the simulations and gathering statistics is as follows. We perform precursor simulations to generate a realistic poured-bed as described in \S\ref{sec:bed_formation}. Then, we carry out simulations initialized from quiescent flow. We run the simulations for several periods until the flow reaches a periodic state and loses the effect of initial conditions. This takes about 2 periods. After this, we run the simulations for additional 8 periods to collect and compute phase-averaged statistics. We ensure that the statistics are converged by confirming that adding data from additional periods does not change the phase-averaged statics. 

{\color{revision3}
While the computational cost of these Euler-Lagrange simulations is high, they remain nevertheless achievable with today's supercomputing resources. Each case requires about 400,000 CPU-hours on AMD EPYC 7742 CPUs. This is equivalent to about 15 day run time on 1152 CPUs. The total cost is 1.2M CPU-hours for the three cases with sediment beds.

In contrast, the cost of doing PR-DNS of these cases largely exceeds computing resources  afford to most academic researchers, if not completely intractable. Taking a typical discretization of 16 grid points per diameter ($\Delta x = d_p/16$) \citep{kidanemariamInterfaceresolvedDirectNumerical2014,mazzuoliDirectNumericalSimulations2019,kasbaouiHighfidelityMethodologyParticleresolved2025},   PR-DNS requires 512 times more grid points than an Euler-Lagrange simulation of the same case. Thus, we estimate the computational cost to be 614.4M CPU-hours to complete all three cases. This puts the simulation run time at about 21 years per case on 1152 CPUs.
}

}

\subsection{Bed formation and bed-fluid interface}\label{sec:bed_formation}
\begin{figure}\centering
	\begin{subfigure}[b]{\linewidth}\centering
		\includegraphics[width=4.0in]{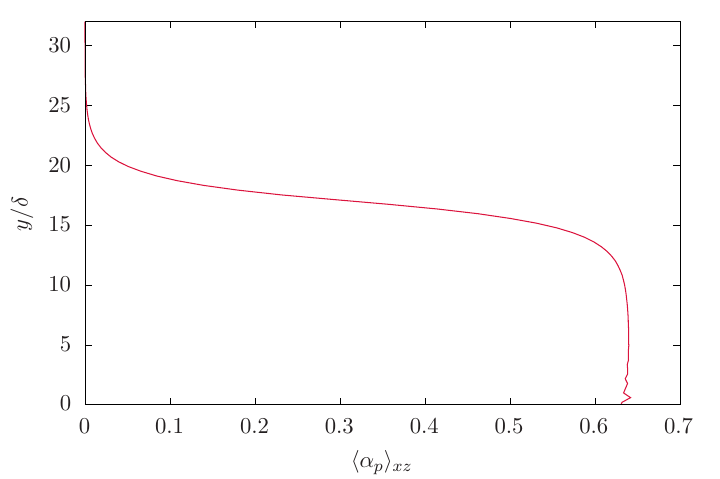}
		\caption{}
		\label{fig:VFP_Profile}
	\end{subfigure}
	\begin{subfigure}[b]{\linewidth}\centering
    	\includegraphics[width=4in,trim={0ex 0ex 0ex 5ex},clip]{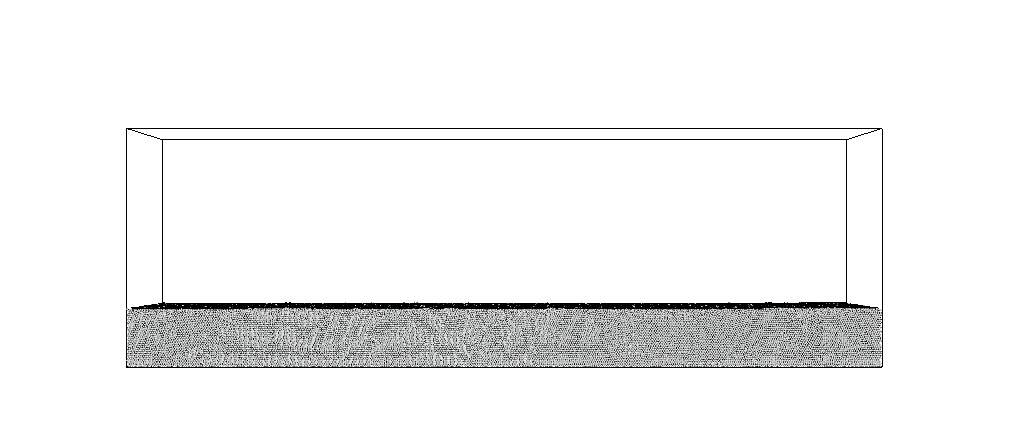}
		\caption{}
		\label{fig:Bed_Interface_Definition}
	\end{subfigure}
	\begin{subfigure}[b]{\linewidth}\centering
    	\includegraphics[width=4in,trim={0ex 0ex 0ex 40ex},clip]{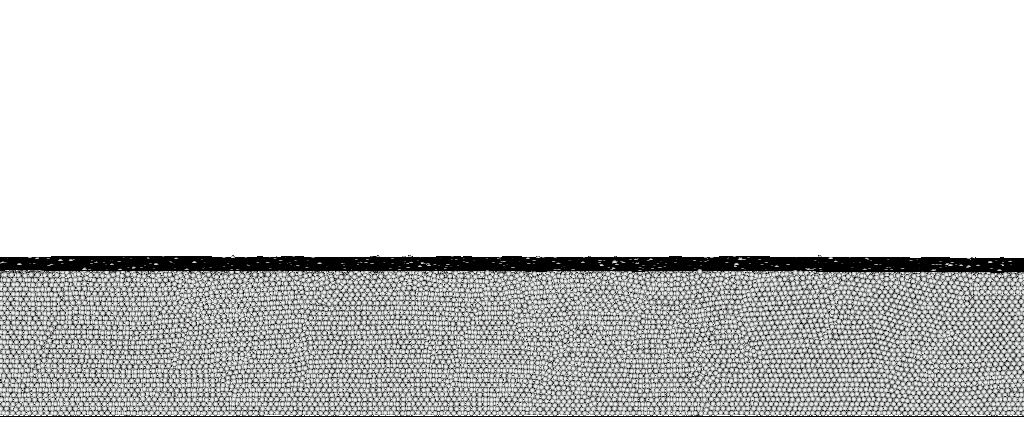}
		\caption{}
		\label{fig:Bed_Interface_Definition_zoom}
	\end{subfigure}
    \caption{The particle bed is initialized by letting particles settle onto the bottom wall. (a) This procedure results in a volume fraction profile that is consistent with that of a poured bed. \textcolor{revision2}{(b,c)} The isosurface $\alpha_p=0.2$ represents a good indicator of the location of the bed-fluid interface.}
\end{figure}

To form the sediment bed, \textcolor{revision}{we perform precursor simulations that serve to generate a realistic bed volume fraction that matches the volume fraction of a poured bed, about $63\%$  \citep{scottDensityRandomClose1969}}. In these runs, the oscillatory forcing is turned off and the particles are initially uniformly distributed towards the middle of the domain at an average volume fraction of 40\% and with small random velocity fluctuations. We then integrate the governing equations (\ref{eq:continuity}), (\ref{eq:momentum}), (\ref{eq:part_pos}), (\ref{eq:lpt_1}), and (\ref{eq:lpt_2}) until the particles fully settle down. Particle-particle collisions and fluid-mediated particle-particle interactions lead to the formation of the poured bed in figure \ref{fig:configuration}. 

Figure \ref{fig:VFP_Profile} shows the average particle volume fraction $\langle \alpha_p\rangle_{xz}$ profile as a function of the wall normal distance. Note that here and onward, the notation $\langle \cdot\rangle_{xz}$ refers to ensemble and spatial averaging over the streamwise ($x$) and spanwise ($z$) directions. As anticipated, the volume fraction within the bed matches the random poured packing  \citep{scottDensityRandomClose1969}. It smoothly transitions to zero away from the bed. Further, we conduct the simulations with particle beds that are sufficiently deep to ensure that  the interaction between the particle bed and the turbulent flow above is captured without interference from the bottom boundary. In the present study, the sediment bed is thick by about 22 particle diameters, which corresponds to about \textcolor{revision}{$\sim 16.7 \delta$.}

At this point, we must address the way we define the bed-fluid interface. We follow the approach of \citet{kidanemariamInterfaceresolvedDirectNumerical2014}, where we define the bed-fluid interface using an isosurface of the particle volume fraction $\alpha_{p}=\alpha_{p,b}<0.63$. \textcolor{revision3}{This is also similar to the experimental approach of \citet{aussillousInvestigationMobileGranular2013}, where black white thresholding of sideview frames of the bed are used to detect the bed interface. However, it is important to recognize that the choice of the isosurface $\alpha_{p,b}$ demarcating the bed-fluid interface is somewhat arbitrary since the computation of the volume fraction field $\alpha_{p}$ depends on numerical choices. For example, the shape and size of the filter kernel used to compute $\alpha_{p}$ control the width of the transition region in figure \ref{fig:VFP_Profile}. With the filtering described in \S\ref{sec:governing_equations}, the isosurface $\alpha_{p,b}=0.2$ provides a good indicator of the approximate location of the bed-fluid interface. We determine this by verifying that this surface lies right on top of the particles as shown in figures  \ref{fig:Bed_Interface_Definition}, \ref{fig:Bed_Interface_Definition_zoom}.}

\section{Characteristics of an oscillatory boundary layer over a cohesionless particle bed}\label{sec:particle_bed}

\textcolor{revision}{Before proceeding further, we refer the reader to appendix \ref{sec:appendix_SP} for a review of the flow features observed when the oscillatory boundary layer develops over a smooth and impermeable wall at  $\Rey_\delta= 200$, 400, and 800. These characteristics provide a benchmark for comparison in what follows. Having reviewed these dynamics, we now analyze the changes that occur when the oscillatory boundary layer develops over a cohesionless particle bed.}
 
\subsection{\color{revision}Overview of the dynamics}\label{sec:overview}

{\color{revision}
\begin{figure} \centering
    \includegraphics[width=4in]{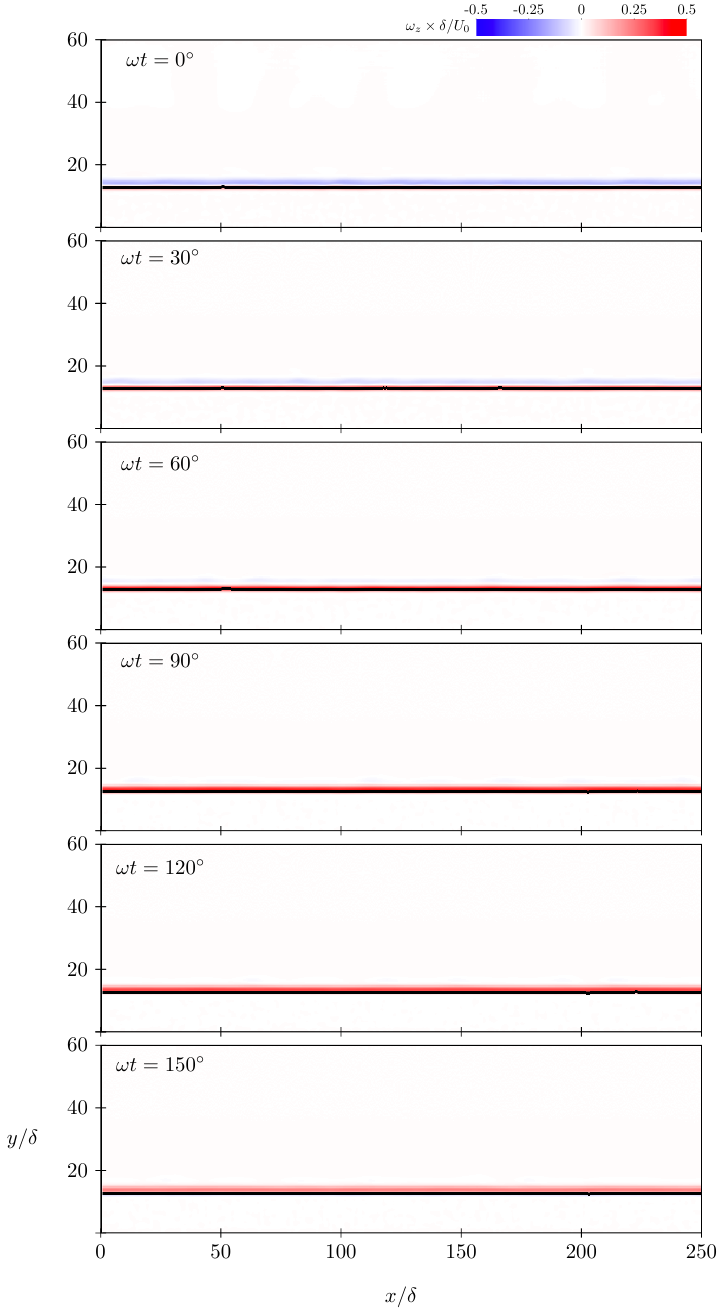}   
    \caption{\color{revision} Zoomed-in view of the instantaneous spanwise vorticity and bed-fluid interface (solid line) at $\Rey_{\delta}=200$. Small ripples in the bedform cause \textcolor{revision2}{flow disturbances} and fluctuations associated with the disturbed laminar regime.}
    \label{fig:Low_Reynolds_Vorticity_200}
\end{figure}

The presence of a sediment bed leads to notable \textcolor{revision2}{flow disturbances}, even at low $\Rey_\delta$ for which DNS of oscillatory boundary layers over smooth and impermeable walls show flow fields devoid of any fluctuations. At $\Rey_\delta=200$, small \textcolor{revision2}{imperfections} in the bed-fluid interface \textcolor{revision2}{are} responsible for \textcolor{revision2}{flow disturbances.} This is shown in figure \ref{fig:Low_Reynolds_Vorticity_200} depicting the instantaneous spanwise vorticity in a wall normal plane at different phases of the cycle. To highlight the bedform, figure \ref{fig:Low_Reynolds_Vorticity_200} also shows the volume fraction contour $\alpha_p=\alpha_{p,b}=0.2$ that demarcates the sediment bed-fluid interface.  The small \textcolor{revision2}{imperfections} in the bed-fluid interface \textcolor{revision2}{are} the result of the initial bed generation as described in \S\ref{sec:bed_formation}. At $\Rey_\delta=200$, the bed shear stress is too low to induce any significant motion of the particles. Thus, the initial bedform persists throughout the simulation. The resulting flow fluctuations are reminiscent of the fluctuations described by \cite{vittoriDirectSimulationTransition1998} in the disturbed laminar regime, where the bottom wall has small waviness. Since a smooth and impermeable wall does not generate such fluctuations at $\Rey_\delta=200$ (see appendix \ref{sec:appendix_SP}), this suggests that \textcolor{revision2}{flow disturbances} induced by asperities in the bed-fluid interface is the driving mechanics \textcolor{revision2}{at this Reynolds number.}

\begin{figure}\centering  
    \includegraphics[width=4in]{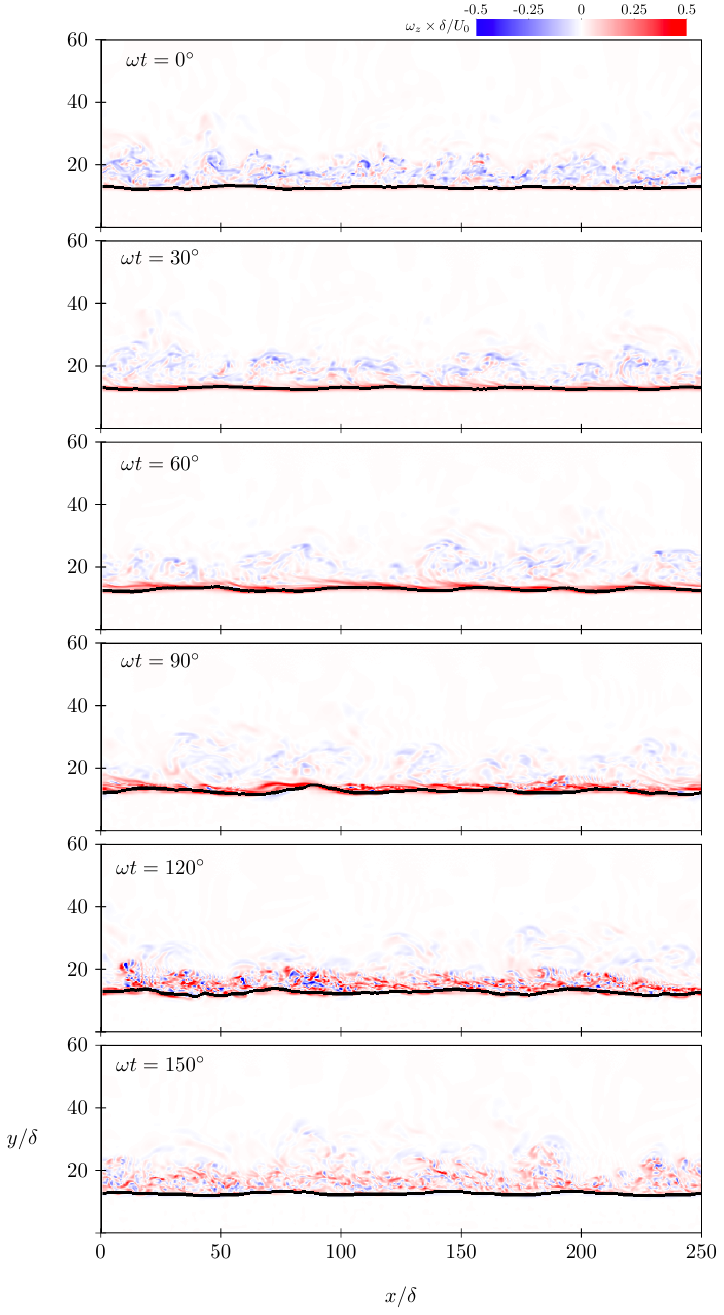}   
    \caption{\color{revision} Zoomed-in view of the instantaneous spanwise vorticity and bed-fluid interface (solid line) at $\Rey_{\delta}=400$. Increasing Reynolds number leads to greater \textcolor{revision2}{flow disturbances} and dynamically-evolving bed-fluid interface.}
    \label{fig:Low_Reynolds_Vorticity_400}
\end{figure}

At $\Rey_\delta=400$, the particles in the bed's top layers become mobile. This leads to a dynamically evolving bed-fluid interface and greater \textcolor{revision2}{flow disturbances}, as shown in figure \ref{fig:Low_Reynolds_Vorticity_400}. \textcolor{revision2}{Flow disturbances are} strongest around phases $90^{\circ}$ and $120^{\circ}$, i.e., from the maximum fluid velocity, and into the decelerating portion of the period. The vortex structures observed at these phases show a chaotic behavior, whereby larger structures spin off and break down into smaller ones. However, the range of scales is limited compared to what may be expected for a fully turbulent flow. The bed-fluid interface, marked by the black line in figure \ref{fig:Low_Reynolds_Vorticity_400}, changes dynamically with phase. This is due to particles being transported in the top layers of the bed, which couples the bedform dynamics to those of the flow over it. 

\begin{figure}	\centering
    \includegraphics[width=4in]{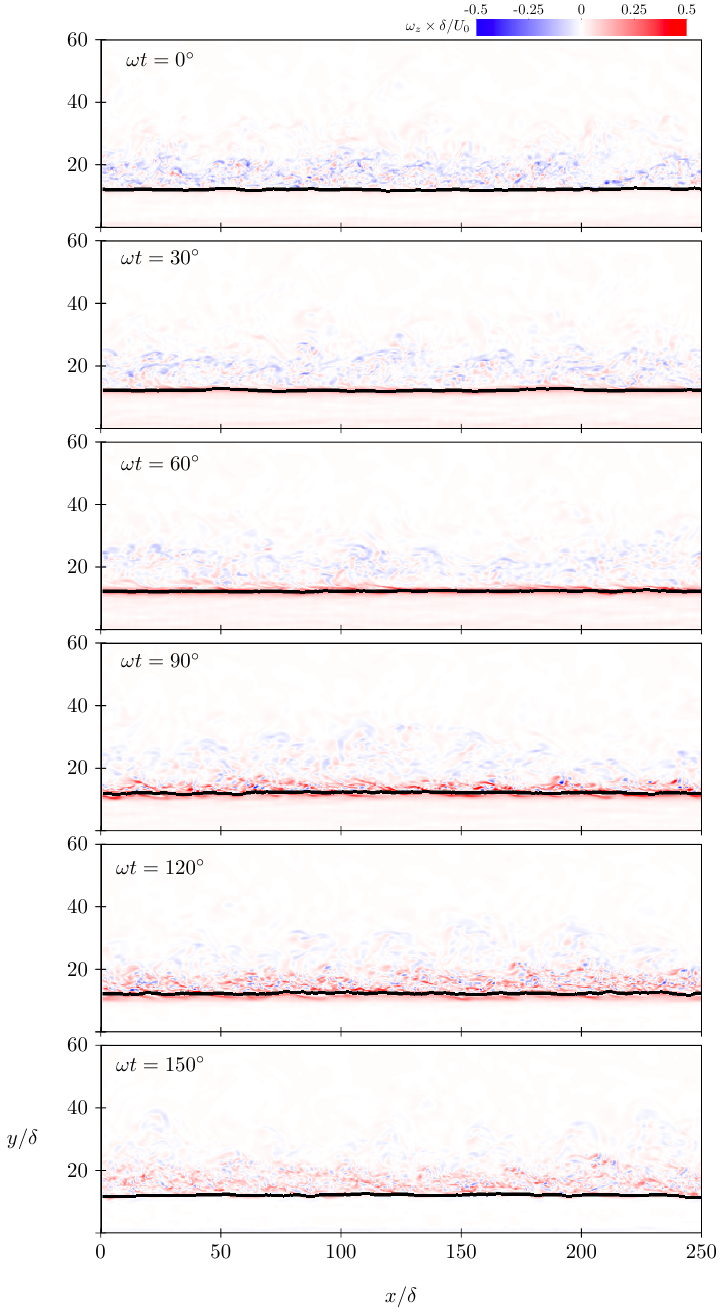}   
    \caption{\color{revision} Zoomed-in view of the instantaneous spanwise vorticity and bed-fluid interface (solid line) at $\Rey_{\delta}=800$. The bedform shifts into \textcolor{revision}{ripples} at various phases. The shedding vortices create a large range of scales. The eddies penetrate the bed interface.}
    \label{fig:High_Reynolds_Vorticity}
\end{figure}

We also note that bed permeability is significant at $\Rey_\delta=400$. Whereas the extent of the flow intrusion below the bed-fluid interface is on the order of one Stokes thickness $\delta$ at $\Rey_\delta=200$, the vortices generated at $\Rey_\delta=400$ penetrate down by up to $4\delta$, judging from figures \ref{fig:Low_Reynolds_Vorticity_200} and \ref{fig:Low_Reynolds_Vorticity_400}. Owing to the bed permeability, these vortices push fluid into and out of the bed. This plays an important role in the dynamic evolution of the bed-fluid interface, as flow exiting the bed exerts an upward drag force on the particles that helps suspend or set into motion particles in the bed's top layers \citep{jewelEffectSeepageFlow2019}.

At $\Rey_\delta = 800$, figure \ref{fig:High_Reynolds_Vorticity} shows significant increase in flow disturbances over the bed, bed-fluid interface corrugations, and flow intrusion within the bed. The fluctuations' intensity and spatial extent exceed largely those due to intermittent turbulence in the case of an oscillatory boundary layer over a smooth and impermeable wall. The flow intrusion within the particle bed is also much greater at $\Rey_\delta=800$ compared to $\Rey_\delta=400$. This likely contributes to the increased corrugation of the bed-fluid interface at this Reynolds number. Further, the bed-fluid interface is most corrugated around phases $60^{\circ}$, $90^{\circ}$, and $120^{\circ}$, which correspond to the phases with largest flow intrusion.
}

\subsection{\color{revision}Fluid Statistics}\label{sec:fluid_stat}

{\color{revision}
Having established the qualitative dynamics of these flows in \S\ref{sec:overview}, we now characterize the fluid phase with quantitative measures.

Figure \ref{fig:velocity_statistics_Re200}, \ref{fig:velocity_statistics_Re400}, and \ref{fig:velocity_statistics_Re800} shows vertical profiles of  phase-averaged mean streamwise fluid velocity $\langle u_f\rangle_{xz}$ from the cases with particle bed at $\Rey_\delta=200$, 400, and 800. To better appreciate the change caused by the particle bed, we also report data from the companion runs with a bottom smooth and impermeable wall discussed in appendix \ref{sec:appendix_SP}. In this figure, $y_b$ denotes the average bed height. We determine the latter by  computing the average $y$ location of the isovolume $\alpha=\alpha_b=0.2$, which represents the bed-fluid interface.

At $\Rey_\delta=200$, the average streamwise fluid velocity from the cases with a particle bed and smooth impermeable wall are sensibly close and follow the laminar Stokes solution. The notable differences include marginally thicker boundary layer, a significant slip velocity $u_{f,I}$ at the bed-fluid interface $y=y_b$, which reaches up to \textcolor{revision3}{$u_{f,I}\simeq 0.1\,U_0$ at phase $60^\circ$}, and interstitial flow that decays quickly within $1.5\delta$ of depth. These features are characteristic of permeable interfaces \citep{voermansVariationFlowTurbulence2017}, although their net effect on the average streamwise fluid velocity profiles at $\Rey_\delta=200$ is limited.

With increasing $\Rey_\delta$, the differences between cases with particle bed and cases with a smooth and impermeable wall accentuate as effects due to bed permeability effects increase. Most notably, we note the increase of the boundary layer thickness, interfacial slip velocity, and depth of the interstitial flow. At $\Rey_\delta=400$, the interfacial slip velocity peaks at \textcolor{revision3}{$u_{f,I}\simeq 0.17\,U_0$ at phase $60^\circ$, while the flow extends below the bed-fluid interface by up to $9\delta$}. At $\Rey_\delta=800$, the maximum slip velocity  increases to \textcolor{revision3}{$u_{f,I}\simeq 0.52\, U_0$ and the flow extends below the bed-fluid interface by up to $14\, \delta$.}

\begin{figure}\centering
    \begin{subfigure}{0.49\linewidth}\centering
    \includegraphics[width=\linewidth]{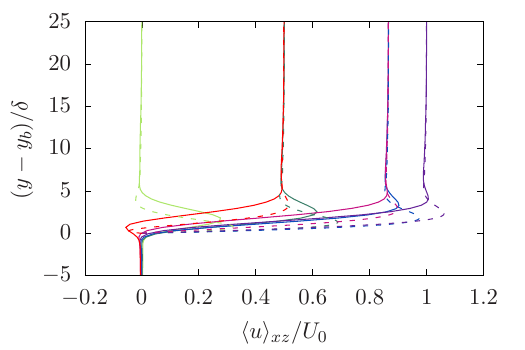}
    \caption{$\Rey_\delta = 200$}
    \label{fig:velocity_statistics_Re200}
    \end{subfigure}
    \begin{subfigure}{0.49\linewidth}\centering
    \includegraphics[width=\linewidth]{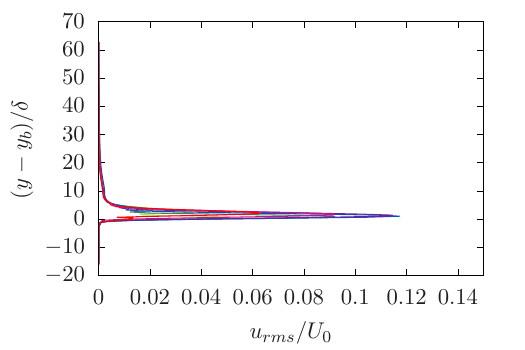}
    \caption{$\Rey_\delta = 200$}
    \label{fig:velocity_statistics_Re200_rms}
    \end{subfigure}
    \begin{subfigure}{0.49\linewidth}\centering
    \includegraphics[width=\linewidth]{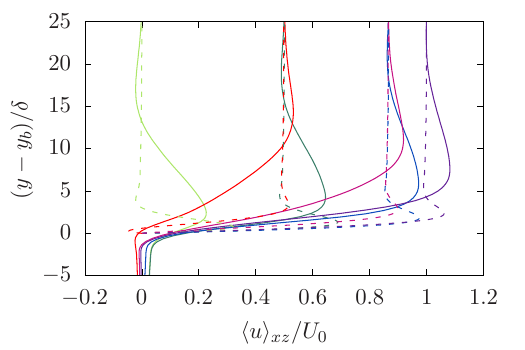}
    \caption{$\Rey_\delta = 400$}
    \label{fig:velocity_statistics_Re400}
    \end{subfigure}
    \begin{subfigure}{0.49\linewidth}\centering
    \includegraphics[width=\linewidth]{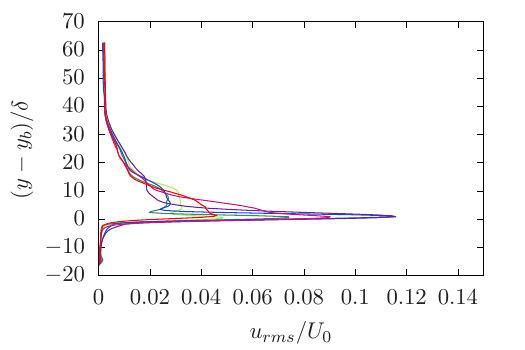}
    \caption{$\Rey_\delta = 400$}
    \label{fig:velocity_statistics_Re400_rms}
    \end{subfigure}
    \begin{subfigure}{0.49\linewidth}\centering
    \includegraphics[width=\linewidth]{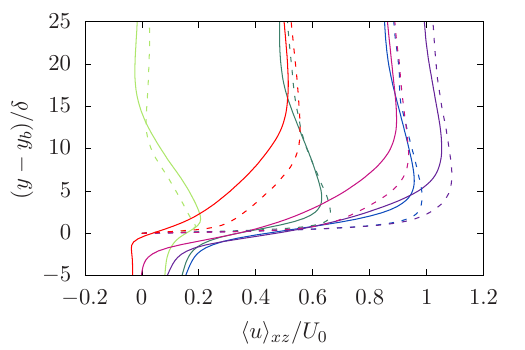}
    \caption{$\Rey_\delta = 800$}
    \label{fig:velocity_statistics_Re800}
    \end{subfigure}
    \begin{subfigure}{0.49\linewidth}\centering
    \includegraphics[width=\linewidth]{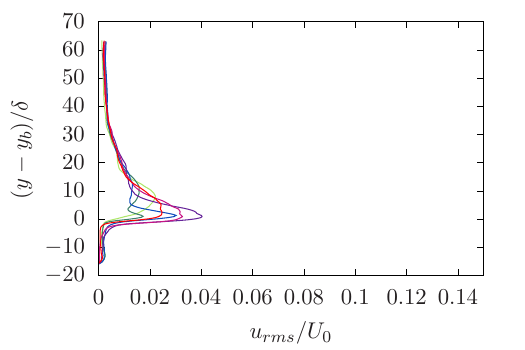}
    \caption{$\Rey_\delta = 800$}
    \label{fig:velocity_statistics_Re800_rms}
    \end{subfigure}
    \caption{Statistics of the phase-averaged mean streamwise velocity at (a,b) $\Rey_\delta=200$, (c,d) $\Rey_\delta=400$, and (e,f) $\Rey_\delta=800$. \textcolor{revision}{The lines correspond to phases $\omega t =0$ (\textcolor{gpltphase0}{\LineStyleSolid}), $\omega t =30$ (\textcolor{gpltphase30}{\LineStyleSolid}), $\omega t =60$ (\textcolor{gpltphase60}{\LineStyleSolid}), $\omega t =90$ (\textcolor{gpltphase90}{\LineStyleSolid}), $\omega t =120$ (\textcolor{gpltphase120}{\LineStyleSolid}), and $\omega t =150$ (\textcolor{gpltphase150}{\LineStyleSolid}).} \textcolor{revision}{The dashed lines correspond to the smooth wall simulations from appendix \ref{sec:appendix_SP}.}}
    \label{fig:velocity_statistics}
\end{figure}

In addition to altering the mean velocity profiles, the presence of a particle bed leads to greater velocity fluctuations than in the smooth wall cases. Figures \ref{fig:velocity_statistics_Re200_rms}, \ref{fig:velocity_statistics_Re400_rms}, and \ref{fig:velocity_statistics_Re800_rms} show the streamwise velocity fluctuations for each Reynolds number. While the root mean square (rms) of the streamwise velocity fluctuations $u_{f,rms}$ in cases with smooth and impermeable wall at $\Rey_\delta = 200$ and 400 are identically zero, we note the existence of significant fluctuations in the cases with particle bed and at matching Reynolds numbers. For $\Rey_\delta = 200$, $u_{f,rms}$ peaks at about \textcolor{revision3}{13\%} of the \textcolor{revision}{velocity} amplitude at about \textcolor{revision3}{$1.4\delta$} above the bed. At $\Rey_\delta = 400$, $u_{f,rms}$ \textcolor{revision3}{the peak is 13\% of the velocity amplitude $U_0$ at phase $90^\circ$ and about  $1.5 \delta$} above the bed. At $\Rey_\delta = 800$, the $u_{f,rms}$ profiles are widest indicating that the flow disturbances extend to about $20\delta$ to $30\delta$ above the bed. The highest fluctuations occur at phase \textcolor{revision3}{$90^\circ$ and reach about 6.9\% of the velocity amplitude.} \textcolor{revision2}{It is important to note that at $\Rey_\delta = 400$ and $800$ velocity fluctuations are no longer homogenous in the streamwise direction due to the waviness of the fluid-bed interface.} \textcolor{revision3}{For $\Rey_\delta = 400$ and $800$, the velocity fluctuations drop to below 0.1\% of the velocity amplitude ($U_0$) at $60\delta$ from the fluid-bed interface.}

 The particle bed leads to a different condition at the fluid-bed interface as compared to a smooth wall. In the smooth wall case, no-slip applies at the wall, while the particle bed is porous, which leads to a slip velocity at the fluid-bed interface. This causes the bed shear stress to drop compared to the smooth wall case. \textcolor{revision3}{Predicting the sediment transport is dependent upon accurate values of the bed shear stress, or nondimensionaly, the Shields number, $\theta$. The Shields number can be estimated a priori using the single phase wall shear stress, i.e., $\tau_w=\mu \left.\frac{\partial u}{\partial y}\right|_{y=0}$.  We denote this Shields number as $\theta_\mathrm{wall}=\tau_w/((\rho_p-\rho_f)g d_p)$. Alternatively, the bed shear stress can be defined using the shear stress conditioned on an isosurface corresponding to the bed-fluid interface $\alpha_{p} = \alpha_{p,b}$ \citep{arollaTransportModelingSedimenting2015}. We evaluate this in two ways. The first way follows the calculation of \citet{arollaTransportModelingSedimenting2015}, that is
 \begin{equation}
 	\tau_b = (\mu +\mu^\star)  \left. \langle \frac{\overline{\partial u}}{\partial y} \rangle \right|_{y_b}=\mu \alpha^{-2.8}  \left. \langle \frac{\overline{\partial u}}{\partial y} \rangle \right|_{y_b}
 \end{equation}}
 \textcolor{revision3}{We denote the Shields number based on the expression above as $\theta_\mathrm{bed}^{(1)}$. Note that this evaluation does not account for the waviness of the bedform. To remediate to this, we carry out an alternate computation of the bed shear stress by interpolating the deviatoric stress tensor to the bed-fluid interface.} \textcolor{revision}{Figure \ref{fig:Re_400_Bed_STL_Vis} shows an example of instantaneous isosurface $\alpha=\alpha_b$ representing the bed-fluid interface. In this second approach, we determine the bed shear $\tau_b$ using}
\begin{equation}
    \tau_b = \left|\left|\frac{1}{A_I}\iint_{S_I} \boldsymbol{n}\cdot\boldsymbol{\tau'}dS\right|\right|
\end {equation}
where $S_I$ represents the bed-fluid interface, $A_I$ is the total interfacial area, $\boldsymbol{n}$ is the normal vector on the isosurface $\alpha_{p}=\alpha_{p,b}$, and $\boldsymbol{\tau'}=\mu [\nabla \boldsymbol{u} +\nabla \boldsymbol{u}^T- (2/3)(\nabla\cdot\boldsymbol{u})\boldsymbol{I}]+\boldsymbol{R_\mu}$ is the deviatoric stress tensor. With the closure of \citet{gibilaroApparentViscosityFluidized2007}, this tensor reads
\begin{equation}
  \boldsymbol{\tau'}=\mu \alpha_f^{-2.8} \left(\nabla \boldsymbol{u} +\nabla \boldsymbol{u}^T- (2/3)(\nabla\cdot\boldsymbol{u})\boldsymbol{I}\right)
\end{equation}
\textcolor{revision3}{We denote the resulting Shields number as $\theta_\mathrm{bed}^{(2)}$.} 

\textcolor{revision3}{Figure \ref{fig:Shields_numb} shows the evolution of $\theta_\mathrm{wall}$, $\theta_\mathrm{bed}^{(1)}$, and $\theta_\mathrm{bed}^{(2)}$ during full cycle. We also report the maximum Shields number obtained with  each method in table \ref{tab:Shields Numbers}. We note close agreement between $\theta_\mathrm{bed}^{(1)}$  and $\theta_\mathrm{bed}^{(2)}$ at all Reynolds numbers. This indicates that the bed waviness does not have a significant effect on the bed shear stress in these cases. Regardless of the method used, the Shields number estimated using the bed shear stress ($\theta_\mathrm{bed}^{(1)}$ and $\theta_\mathrm{bed}^{(2)}$) is much lower than the Shields number based on single-phase wall shear stress $\theta_\mathrm{wall}$, even at $\Rey_\delta=200$. This was also noted by
\citet{kidanemariamInterfaceresolvedDirectNumerical2014} in PR-DNS of laminar channel flow with a sediment bed. They observed a progressive departure of the Shields number from the Poiseuille predictions with increasing fluid flow rate, caused by the departure of the flow profile from a strictly parabolic profile at the bed interface. Likewise, flow intrusion through the sediment bed leads to lower gradients at the bed as shown in figure \ref{fig:velocity_statistics}, which in-turn, leads to lower bed shear stress. }

\begin{figure}\centering
    \hfill\includegraphics[scale=0.75]{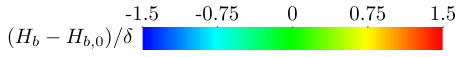}\\
	 \includegraphics[width=4in,trim={40ex 50ex 15ex 70ex},clip]{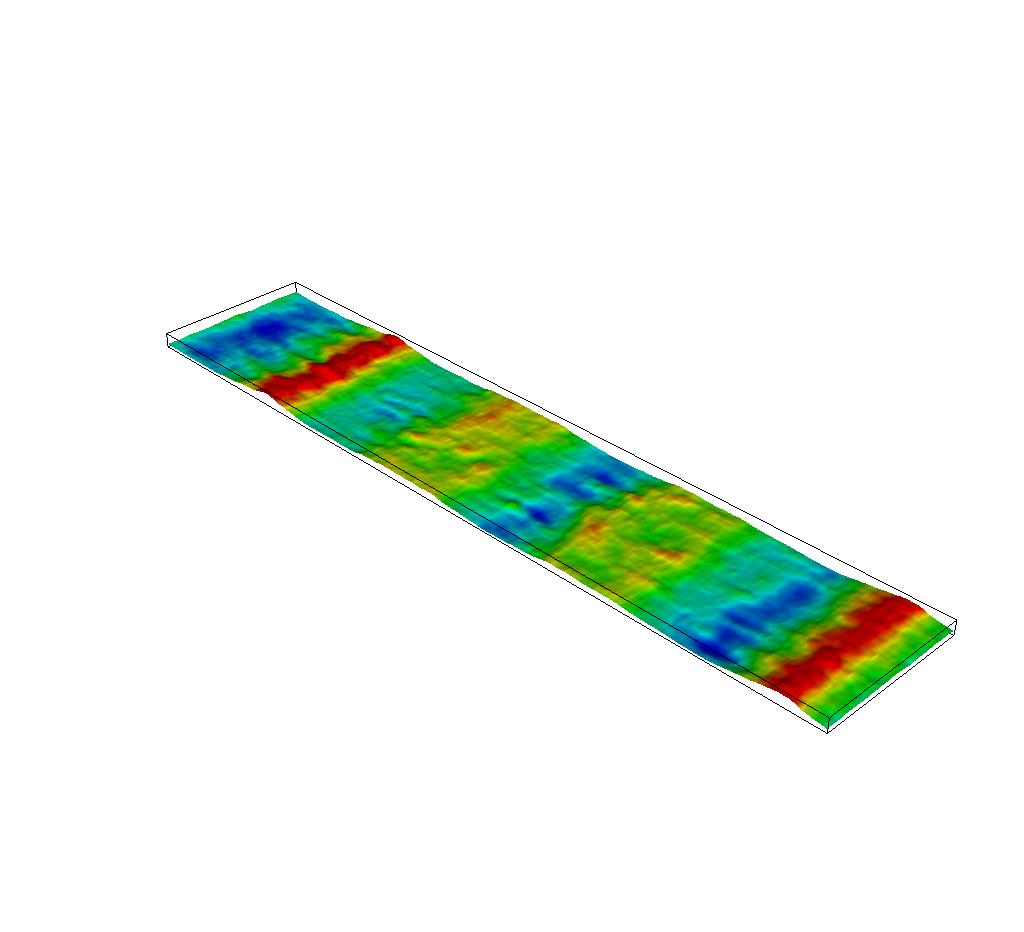}
	\caption{\textcolor{revision}{$\Rey_\delta = 400$} bedform height deviations. Small ripples rise and fall below the average bed height.}
    \label{fig:Re_400_Bed_STL_Vis}
\end{figure}

\begin{figure}
    \centering
    \begin{subfigure}{0.49\linewidth}\centering
    \includegraphics[width=2.75in]{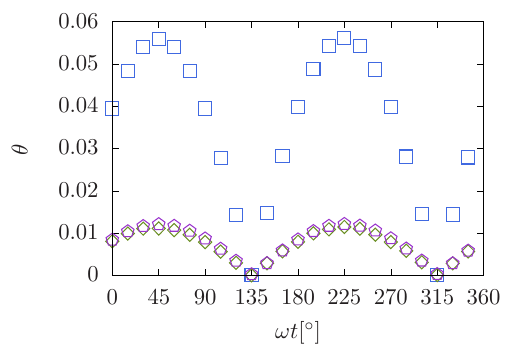}
    \caption{$\Rey_\delta = 200$}
    \end{subfigure}
    \begin{subfigure}{0.49\linewidth}\centering
    \includegraphics[width=2.75in]{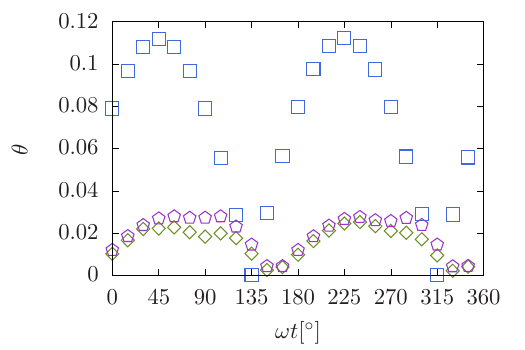}
    \caption{$\Rey_\delta = 400$}
    \end{subfigure}
    \begin{subfigure}{0.49\linewidth}\centering
    \includegraphics[width=2.75in]{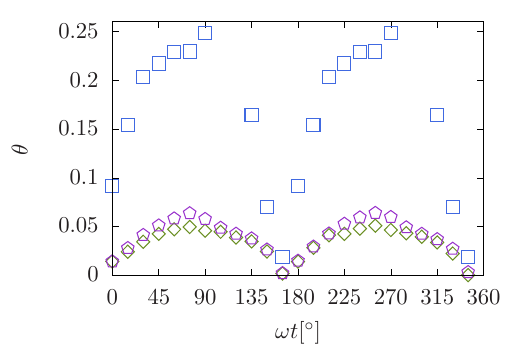}
    \caption{$\Rey_\delta = 800$}
    \end{subfigure}
    \caption{\color{revision3} Variation of the Shields number with the phase. Squares: $\theta_\mathrm{wall}$; Pentagons: $\theta_\mathrm{bed}^{(1)}$; Diamonds: $\theta_\mathrm{bed}^{(2)}$.}
    \label{fig:Shields_numb}
\end{figure}

\begin{table}\color{revision3}
    \begin{tabularx}{\linewidth}{XXXX}
         $\Rey_\delta$ & $\theta_\mathrm{wall}$ & $\theta_\mathrm{bed}^{(1)}$ & $\theta_\mathrm{bed}^{(2)}$ \\[1ex]
         200 & 0.0560 & 0.0296 & 0.0279 \\
         400 & 0.1121 & 0.0330 & 0.0423 \\
         800 & 0.2428 & 0.0504 & 0.0448 
    \end{tabularx}
\caption{\textcolor{revision3}{Maximum values of the Shields number by method of computation.}}
    \label{tab:Shields Numbers}
\end{table}


\subsection{\color{revision}Particle Statistics}
\begin{figure}\centering
    \begin{subfigure}{0.49\linewidth}\centering
    \includegraphics[width=\linewidth,clip]{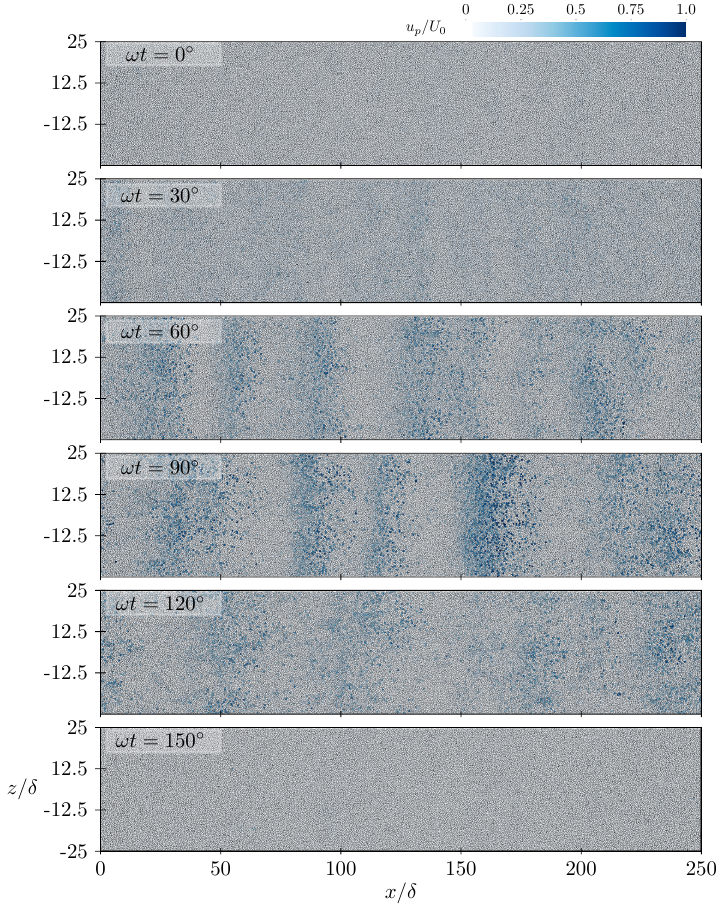}
    \caption{$\Rey_\delta = 400$}
    \label{fig:Particle_Beds_a}
    \end{subfigure}
    \begin{subfigure}{0.49\linewidth}\centering
    \includegraphics[width=\linewidth,clip]{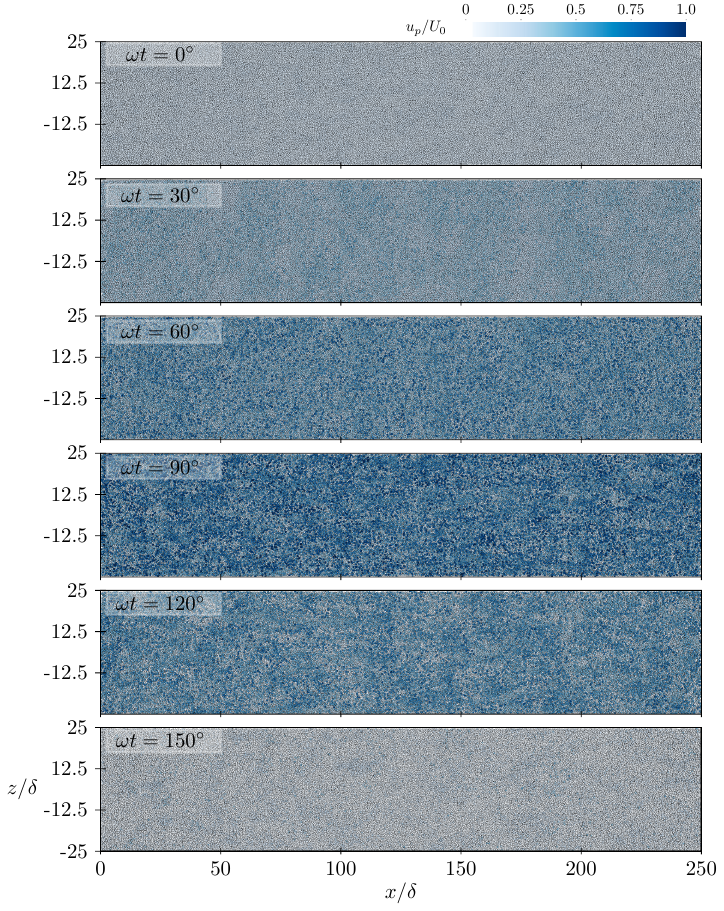}
    \caption{$\Rey_\delta = 800$}	
    \label{fig:Particle_Beds_b}
    \end{subfigure}
        \caption{\color{revision} Top-down view of particle bed at phases $0^\circ$, $30^\circ$, $60^\circ$, $90^\circ$, $120^\circ$, $150^\circ$ degrees for $\Rey_\delta = 400$ and $800$. The particles are colored by their normalized stream-wise velocity. The case at $\Rey_\delta=400$ shows periodic rolling ripples, whereas the case at $\Rey_\delta=800$ \textcolor{revision2}{exhibits a particle suspension layer, and may be evolving towards a new bedform.}}
    \label{fig:Particle_Beds}
\end{figure}
{\color{revision}
Next, we analyze the characteristics of the particulate phase and how these relate to those of the fluid phase.

At $\Rey_\delta=400$, the particle bed is characterized by the periodic particle transport at the ripple crests. This can be seen in figure \ref{fig:Particle_Beds_a} which depicts a top-down view of the particle bed between $\omega t=0^\circ$ and $\omega t=150^\circ$ and where the particles are colored by their normalized streamwise velocities. At phase $0^\circ$, most particles are immobile and within the bed. From there, the rising fluid velocity initiates particle saltation at the ripple crests, which are well visible at phase $60^\circ$. These ripples are quasi-2D and display a wavelength of about \textcolor{revision3}{$\sim 60\delta$.} The rolling ripples continue intensifying until phase $90^\circ$. After that, the decrease in fluid velocity leads to a slow down of the particles until all are immobile again by phase $150^\circ$. This process occurs again, albeit in the reverse direction, between phases $180^\circ$ and \textcolor{revision3}{$360^\circ$}.

With increasing $\Rey_\delta$ to 800, \textcolor{revision2}{the bed may be evolving towards a new bedform with wavelength greater than $L_x$}. As shown in figure \ref{fig:Particle_Beds_b}, the rolling ripples noted at $\Rey_\delta=400$ are no longer present at $\Rey_\delta=800$. Instead, we note the periodic emergence and collapse of a layer of suspended particles over the bed. At phase $0^\circ$, most particles start at rest in the bed. As the flow accelerates, particles in the top layer of the bed start saltating, which can be seen at phase $30^{\circ}$. As the fluid velocity continues increasing, more particles are set in motion. At phase $60^\circ$, we note that a large number of particles are suspended within the fluid column. The \textcolor{revision2}{layer} of suspended particles continues growing through phase $120^{\circ}$, by which point the fluid velocity has begun decreasing. With the continued slowing down of the flow, most particle redeposit in the bed by  phase $150^\circ$, while only few remain suspended.

Note that we do not present similar figures to  \ref{fig:Particle_Beds_a} and  \ref{fig:Particle_Beds_b} for the case at $\Rey_\delta =200$ because particle motion is negligible in that case.

\begin{figure}
 	\centering
 	\begin{subfigure}{0.49\linewidth}
    \includegraphics[width=\linewidth]{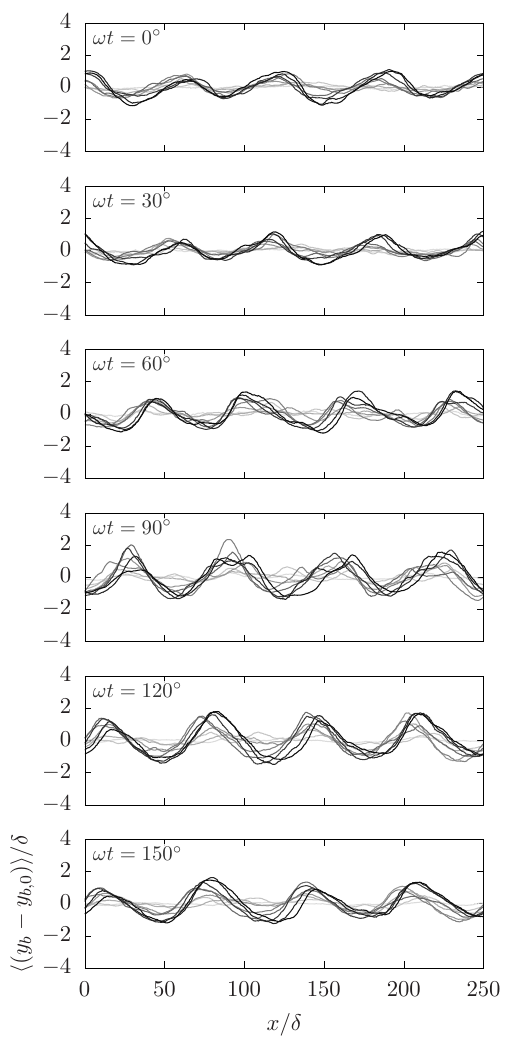}
    \caption{}
    \label{fig:Bed_Interface_Periods_a}
    \end{subfigure}
 	\begin{subfigure}{0.49\linewidth}
    \includegraphics[width=\linewidth]{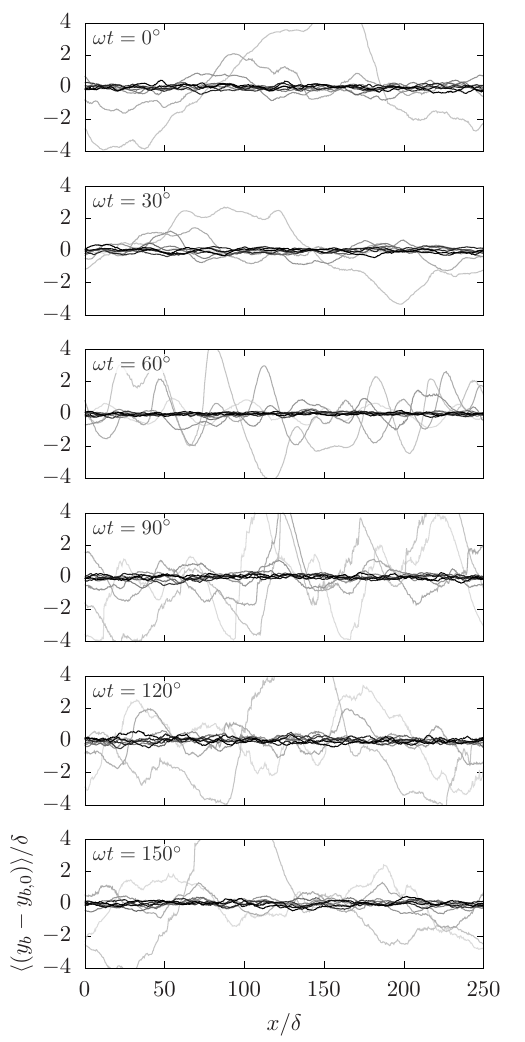}
    \caption{}
    \label{fig:Bed_Interface_Periods_b}
    \end{subfigure}
 	\caption{\color{revision} Bed interface for (a) $\Rey_\delta = 400$ and (b) $\Rey_\delta = 800$. Darker lines indicate later periods. For $\Rey_\delta = 400$, rolling grain ripples emerge and move through domain, but the dominant wavelength does not change. At $\Rey_\delta = 800$, the bed height becomes highly corrugated, and the bed fluid interface breaks down.}
 	\label{fig:Bed_Interface_Periods}
 \end{figure}
 
Figure \ref{fig:Bed_Interface_Periods} shows the bed surface interface averaged over the spanwise direction, for $\Rey_\delta = 400$ and 800. The different lines correspond to the interface location from periods 2 to 10. At $\Rey_\delta = 400$, the bed-fluid interface in figure \ref{fig:Bed_Interface_Periods_a} develops into a clear bedform as the simulation progress. The wavelength associated with the ripples is \textcolor{revision3}{$\lambda/\delta = 62.6$.} The typical height of the ripples, measured from depression to peak, is about $4\delta$ at phase $90^\circ$. While sediment particles move along the bed at significant velocities, as seen from figure \ref{fig:Particle_Beds_a}, the bedform evolves on a much slower periodic time scale. At $\Rey_\delta = 800$, the bed surface displays a range of small scale corrugations. However, the large scale ripples disappear. Figure \ref{fig:Bed_Interface_Periods_b} shows that the bed height deviations drop below $2\delta$. \textcolor{revision2}{This is because particle transport takes place primarily in a suspension layer.}

Next, we report the vertical profiles of normalized particle \textcolor{revision3}{velocity} in figures \ref{fig:part_momentum_Re400} and \ref{fig:part_momentum_Re800}. At $\Rey_\delta=400$, the normalized particle \textcolor{revision3}{velocity is small, as $\langle\alpha_p u_p\rangle_{xz}/U_0<0.025$ for all phases.} This indicates that the rolling ripples move at a velocity much smaller than the fluid velocity amplitude $U_0$. The largest particle \textcolor{revision3}{velocity occurs at phase $30^\circ$}, in agreement with the qualitative observations from figure \ref{fig:Particle_Beds_a}. We note that the particle \textcolor{revision3}{velocity} is non-zero only in a region of width $\sim 8\delta$ around the average bed heigh $y_b$. The width of this region is controlled by the height of the ripples, noted in figure \ref{fig:Bed_Interface_Periods_a}, and is not due to suspended particles, as the latter are negligible at $\Rey_\delta=400$.
In contrast, the normalized particle \textcolor{revision3}{velocity} reaches significantly higher values and displays wider distribution at $\Rey_\delta=800$. At phase $150^\circ$, the region of non-zero particle momentum is about $18\delta$ thick and peaks at about $0.08\rho_p U_0$. The widening of the profile and the absence of large scale deformations of the bed-fluid interface indicate that the amount of suspended particles is significantly larger at $\Rey_\delta=800$ compared to $\Rey_\delta=400$. This is also in agreement with the qualitative observations from figure \ref{fig:Particle_Beds_b}.

With regards to the computational aspects, we note that the particle \textcolor{revision3}{velocity} is negligible about $\sim 10\delta$ below the bed-fluid interface $y_b$ at all phases and $\Rey_\delta$ considered. This shows that the particle bed in our simulations, which has an initial height $H_b=16.7\delta$, is sufficiently tall to capture the dynamics near the bed-fluid interface without interference from boundary conditions at the bottom of the domain.

  \begin{figure}
 	\centering
 	\begin{subfigure}{0.49\linewidth}\centering
    \includegraphics[width=2.75in]{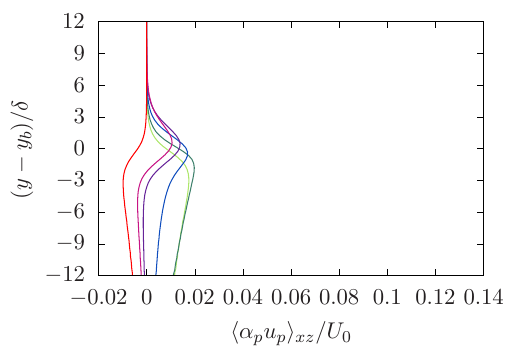}
    \caption{}
    \label{fig:part_momentum_Re400}
    \end{subfigure}
    \begin{subfigure}{0.49\linewidth}\centering
    \includegraphics[width=2.75in]{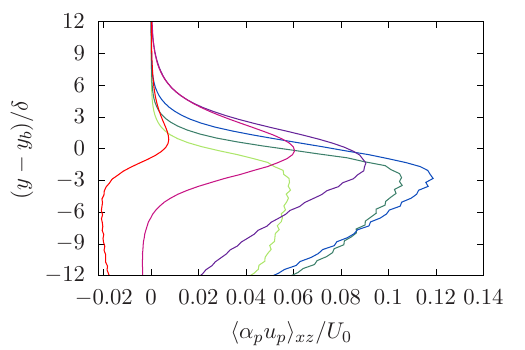}
    \caption{}
    \label{fig:part_momentum_Re800}
    \end{subfigure}
  	\begin{subfigure}{0.49\linewidth}\centering
    \includegraphics[width=2.75in]{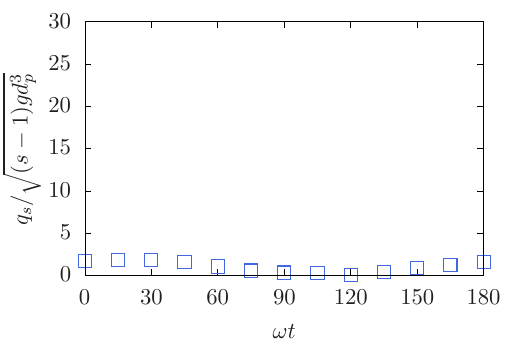}
    \caption{}
    \label{fig:part_mass_flux_Re400}
    \end{subfigure}
    \begin{subfigure}{0.49\linewidth}\centering
    \includegraphics[width=2.75in]{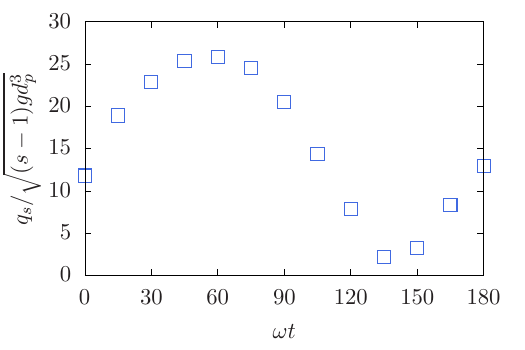}
    \caption{}
    \label{fig:part_mass_flux_Re800}
    \end{subfigure}
 	\caption{
 	Normalized particle momentum profiles and mass fluxes  at $\Rey_\delta=400$ (a,c) and $\Rey_\delta=800$ (b,d). The lines in (a,b) correspond to phases: ($\omega t=0$,\textcolor{gpltphase0}{\LineStyleSolid}), ($\omega t=30$,\textcolor{gpltphase30}{\LineStyleSolid}), ($\omega t=60$,\textcolor{gpltphase60}{\LineStyleSolid}),($\omega t=90$, \textcolor{gpltphase90}{\LineStyleSolid}), ($\omega t=120$,\textcolor{gpltphase120}{\LineStyleSolid}), ($\omega t=150$,\textcolor{gpltphase150}{\LineStyleSolid}). Significant particle momentum is seen near the bed interface, indicating particle motion at around the bed interface.}
 	\label{fig:part_momentum}
 \end{figure}
 
 Figures \ref{fig:part_mass_flux_Re400} and \ref{fig:part_mass_flux_Re800} show the evolution of the normalized sediment flow rate $q_s$ over a half period, for $\Rey_\delta = 400$ and $800$. We compute $q_s$ by integrating the particle velocity profiles over the vertical direction, i.e., 
 \begin{equation}
 	q_s = \int_{0}^{L_y} \langle  \alpha_p u_p \rangle_{xz} dy.
 \end{equation}
At $\Rey_\delta = 400$, the sediment volumetric flux peaks at \textcolor{revision3}{$1.76$ at phase $\omega t = 30^\circ$.} Notably, the sediment transport is not symmetric for the accelerating and decelerating portions of the periods. At $\Rey_\delta = 800$, the sediment flux is more than \textcolor{revision3}{12} times greater than at $\Rey_\delta=400$. \textcolor{revision3}{The maximum normalized flux is $25.8$ at phase $60^\circ$.} Here too, we note an asymmetry between the acceleration and deceleration portions of the period.
}

{\color{revision}
\section{Discussion and conclusions}\label{sec:conclusion}
In this study, we analyze data obtained from Euler-Lagrange simulations of an oscillatory boundary layer developing over a bed of collisional and freely moving sediment grains with density ratio $\rho_p/\rho_f=2.65$ and Galileo number $\mathrm{Ga}=51.9$. In these simulations, we vary the velocity amplitude to yield Reynolds numbers $\Rey_\delta=200$ to 800. The maximum Shields number \textcolor{revision3}{based on the smooth wall case} also varies from  \textcolor{revision3}{$5.60\,10^{-2}$} to \textcolor{revision3}{$2.43\,10^{-1}$} and the Keulegan-Carpenter number varies from $134.5$ to $538.0$. Companion simulations of an oscillatory boundary layer over a smooth and impermeable wall show that the boundary layer follows the laminar Stokes solution at $\Rey_\delta=200$ and $\Rey_\delta=400$ and displays intermittent turbulence at $\Rey_\delta=800$. However, the presence of a mobile bed alters the flow characteristics and leads to a coupling of the dynamics of the oscillatory boundary layer with those of the sediment bed.

The coupled dynamics arise from two fundamental mechanisms. The first one relates to the bed permeability. This is due to the porosity of the bed which allows the flow to penetrate within the sediment layers. The extent of the flow penetration depends on the Reynolds number. In the case at $\Rey_\delta=200$, the flow penetration is on the order of the Stokes thickness $\delta$. With increasing $\Rey_\delta$, the flow penetration increases to about \textcolor{revision3}{$9\delta$ and $14\delta$} at $\Rey_\delta=400$ and $\Rey_\delta=800$, respectively. Another consequence of the bed permeability is the emergence of a slip velocity at the bed-fluid interface. \textcolor{revision3}{At phase $30^\circ$ of the cycle, the interfacial slip velocity peaks at about $0.12U_0$ at $\Rey_\delta=200$,  $0.17U_0$ at phase $60^\circ$ for $\Rey_\delta=400$, and $0.52U_0$, at $90^\circ$ for $\Rey_\delta=800$.} This significant slip leads to a thickening of the boundary layer and reduction of the bed shear stress. 

The second fundamental mechanism that couples the dynamics of the particle bed and the oscillatory boundary layer relates to particle motion. In the case at $\Rey_\delta=200$, the particles remain immobile during the entire cycle. This is expected because the maximum Shields number \textcolor{revision3}{based on the shear stress at the bed-fluid interface} in this case \textcolor{revision3}{($\Theta_\mathrm{max}=1.21\, 10^{-2}$)} is below the critical Shields number for incipient motion. Thus, the only changes to the oscillatory boundary layer in this case are those due to permeability. At $\Rey_\delta=400$ and 800, the particles become mobile and lead to a dynamic evolution of the sediment bed. At $\Rey_\delta=400$, the particles form rolling-grain ripples. The latter emerge from small scale corrugations in the bed-fluid interface. These corrugations coarsen very quickly and lead to a bedform with height $4\delta$ and wavelength of about \textcolor{revision3}{$\sim 60\delta$.} Analyzing the vorticity dynamics shows that these ripples cause enhanced vortex shedding in the flow which leads to fluid velocity disturbances similar to those observed in the disturbed laminar regime. At $\Rey_\delta=800$, the \textcolor{revision2}{intermittent turbulence leads to the formation of a particle transport layer.} Between phases $60^\circ$ and $120^\circ$ of the half-cycle, the bed-fluid interface recedes slightly and a layer of suspended particles forms. During the rest of the half-cycle, the suspended particles settle into the bed and become nearly motionless at phase $0^\circ$. Because of their large momentum and their feedback force on the fluid, the suspended particles induce significant fluctuations in the flow. This leads to further widening of the boundary layer and greater fluctuations than is observed in the corresponding particle-free case.

In this work, we have also shown that the dynamics of sediment beds under a wide range of oscillatory and unsteady flow conditions can be predicted at reasonable computational cost using carefully designed Euler-Lagrange simulations. While PR-DNS require little to no models, their large computational cost often makes them too restrictive. In comparison with the PR-DNS of \citet{mazzuoliInterfaceresolvedDirectNumerical2020}, our simulations have domains 100 times larger: about 10 times longer in the streamwise direction, about 4 times wider in the spanswise direction, and about twice taller. This allows us to capture bedforms which was not possible in \citet{mazzuoliInterfaceresolvedDirectNumerical2020}. Further, we were able to simulate 10 cycles for each of our cases, compared to 1 to 4 cycles in \citet{mazzuoliInterfaceresolvedDirectNumerical2020}. The computational cost of these simulations is about 400,000 CPU-hours per case (equivalent to about 15 day run time on 1152 CPUs), which is attainable for many researchers with today's computational resources.
}

\backsection[Acknowledgements]{Computing resources through ACCESS award EES230041 and Research Computing at Arizona State University are gratefully acknowledged.}

\backsection[Funding]{This research received no specific grant from any funding agency, commercial or not-for-profit sectors.}

\backsection[Declaration of interests]{The authors report no conflict of interest.}

\backsection[Author ORCIDs]{
	M. H. Kasbaoui, https://orcid.org/0000-0002-3102-0624;
}

\appendix
\section{Structure of an Oscillatory Boundary Layer over a smooth wall} \label{sec:appendix_SP}
In this appendix, we present a detailed description of the flow characteristics at $\Rey_\delta=200$, 400, and 800, where the bottom boundary is an impermeable smooth wall rather than a sediment bed. The reasons for this are twofold. First, analytical solutions exist for the laminar regime, which allows the validation of the computational approach. Second, these runs serve as benchmark to elucidate the changes to the flow in presence of a sediment bed.

\begin{figure}\centering
    \hfill\includegraphics{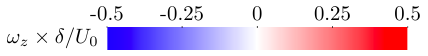}\\
    \begin{subfigure}{0.49\linewidth}\centering \includegraphics[width=2.5in]{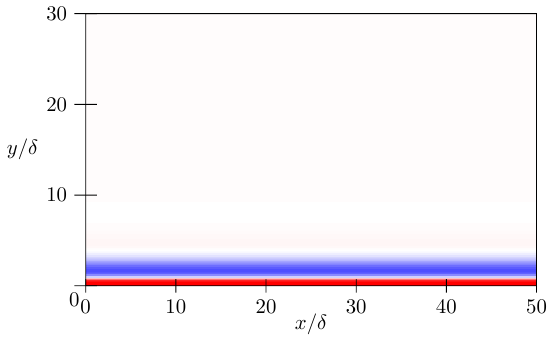} 
    \caption{$\omega t=0^{\circ}$}	
    \end{subfigure}\hfill
    \begin{subfigure}{0.49\linewidth}\centering 
        \includegraphics[width=2.5in]{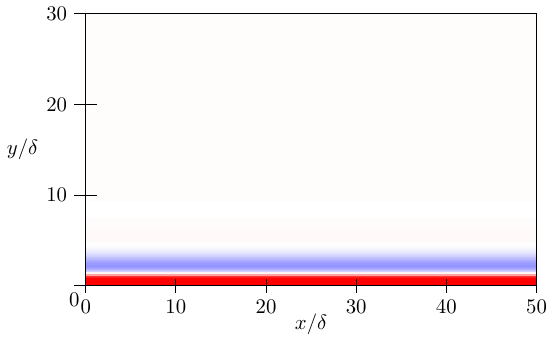}
    \caption{$\omega t=30^{\circ}$}	
    \end{subfigure}\hfill
    \begin{subfigure}{0.49\linewidth}\centering
        \includegraphics[width=2.5in]{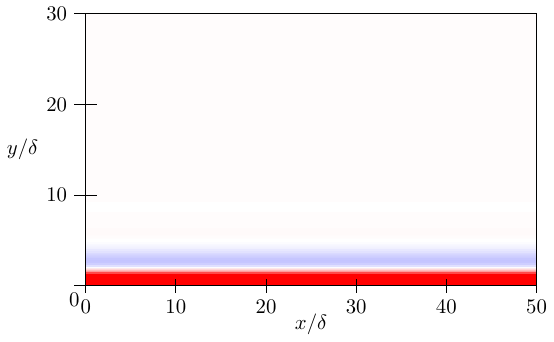}
    \caption{$\omega t=60^{\circ}$}	
    \end{subfigure}\hfill
    \begin{subfigure}{0.49\linewidth}\centering
        \includegraphics[width=2.5in]{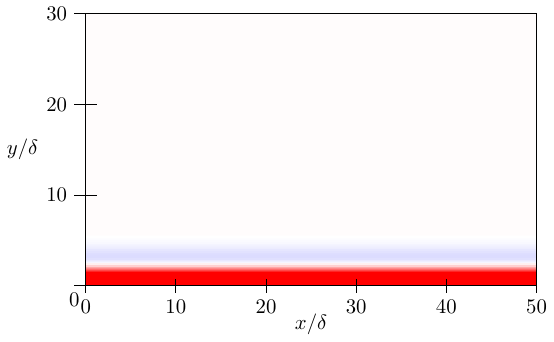}
    \caption{$\omega t=90^{\circ}$}	
    \end{subfigure}\hfill
    \begin{subfigure}{0.49\linewidth}\centering
        \includegraphics[width=2.5in]{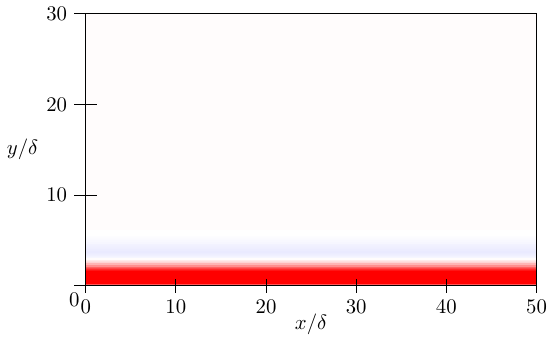}
    \caption{$\omega t=120^{\circ}$}	
    \end{subfigure}\hfill
    \begin{subfigure}{0.49\linewidth}\centering
        \includegraphics[width=2.5in]{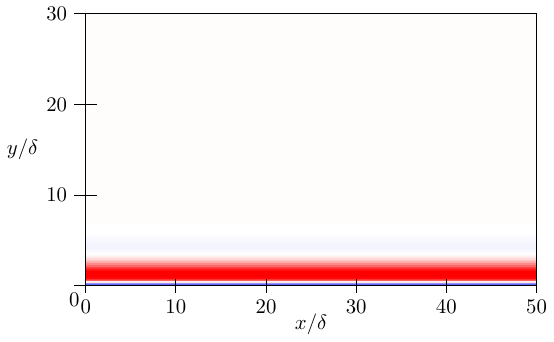}
    \caption{$\omega t=150^{\circ}$}	
    \end{subfigure}\hfill
    \caption{Normalized spanwise vorticity fields at $\Rey_{\delta}=400$ for a smooth, impermeable wall, at phases 0, 30, 60, 90, 120 and 150 degrees. The vorticity is arranged in laminae at all phases.}
    \label{fig:SP_OBL_Vorticity_400}
\end{figure}

\begin{figure}\centering
	\includegraphics[width=4in]{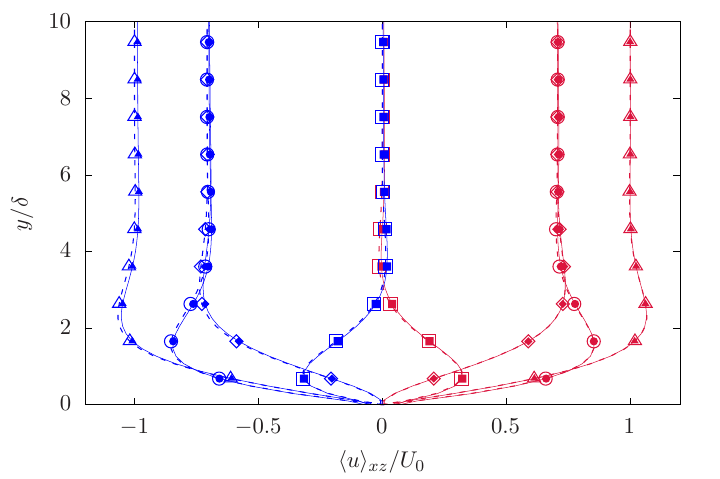}
    \caption{Normalized stream-wise velocity for the case of an OBL over an impermeable smooth wall at $\Rey_\delta=400$. \textcolor{revision}{Filled symbols correspond to the Stokes solution, while open symbols correspond to smooth wall simulations. To differentiate the positive and negative portions of the period, we plot the positive half cycle in red and the negative half cycle in blue.} The symbols indicate the phase: (\textcolor{red}{$\square$}\textcolor{black},$\omega t = 0^{\circ}$, $180^{\circ}$), (\textcolor{red}{$\circ$}\textcolor{black},$\omega t = 45^{\circ}$, $225^{\circ}$), (\textcolor{red}{$\triangle$}\textcolor{black},$\omega t = 90^{\circ}$, $270^{\circ}$), (\textcolor{red}{$\diamond$}\textcolor{black},$\omega t = 135^{\circ}$, $315^{\circ}$). The strong agreement between the simulated data and the Stokes solution indicates that the flow is fully laminar in this case.}
    \label{fig:SP_OBL_Mean_u}
\end{figure}
Figure \ref{fig:SP_OBL_Vorticity_400} shows the normalized spanwise vorticity field at phases $\omega t = 0^{\circ}$, $60^{\circ}$, $90^{\circ}$, $120^{\circ},$ and $180^{\circ}$ for the case \textcolor{revision}{$\Rey_\delta = 400$.} The solution \textcolor{revision}{for $\Rey_\delta = 200$} shows similar vorticity structure to the \textcolor{revision}{$\Rey_\delta = 400$ solution} and, thus, is not included here. The vorticity in these low Reynolds number cases, is organized into sheets in the near-wall region. This indicates laminar flow, and so the flow at \textcolor{revision}{$\Rey_\delta = 200, 400$} should obey the Stokes solution,
\begin{eqnarray}
u_{f,x}/U_0 &=& \cos(\omega t)-e^{-y/\delta}\cos({\omega t - y/\delta}),\\
u_{f,y}/U_0	&=& 0.
\end{eqnarray}
To verify this, we compare the Stokes solution to vertical profiles of the phase-averaged fluid velocity from the simulation at $\Rey_\delta=400$ in figure \ref{fig:SP_OBL_Mean_u}. The agreement between the simulated data and the analytical solution is excellent, showing that the flow is indeed laminar at these Reynolds numbers. This also validates the computational approach for an impermeable smooth wall oscillatory boundary layer at low Reynolds numbers.

\begin{figure}
    \centering
    \includegraphics[width=4in]{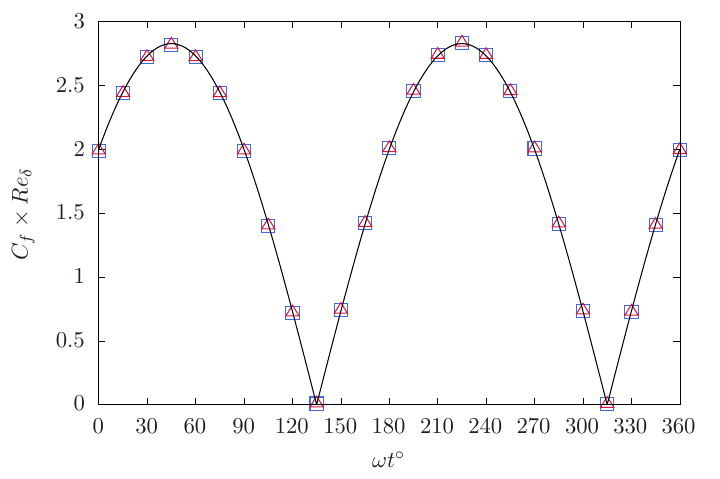}
    \caption{Scaled coefficient of friction over one period at $Re_{\delta} = 200$ and 400, for a smooth, impermeable wall. The black line correspond to the Stokes solution. Symbols correspond to the numerical solution at (\textcolor{gpltBlue}{$\square$}\textcolor{black},$\Rey_{\delta} = 200$) and (\textcolor{gpltRed}{$\triangle$}\textcolor{black},$\Rey_{\delta} = 400$). All cases collapse onto the Stokes solution.}
    \label{fig:sp_cf}
\end{figure}
As an additional comparison, we compute the coefficient of friction $C_f$ defined as,
\begin{equation}
     C_f=\frac{|\tau_w|}{(1/2)\rho_{f} U_0^{2}},
\end{equation}
where $\tau_w$ is the wall shear stress. Figure \ref{fig:sp_cf} shows the variation of the coefficient of friction scaled by $\Rey_{\delta}$ to cancel the Reynolds number dependence of the coefficient of friction. The scaled coefficient is plotted over a period for the cases of an OBL over an impermeable smooth wall at $\Rey_\delta = 200$ and 400, alongside the Stokes solution. Here too, the agreement between numerical and Stokes solution is excellent which further demonstrates that the OBL at these Reynolds number is fully laminar.

\begin{figure}\centering
    \hfill\includegraphics{cb_vorticity.pdf}\\
    \begin{subfigure}{0.49\linewidth}\centering \includegraphics[width=2.5in]{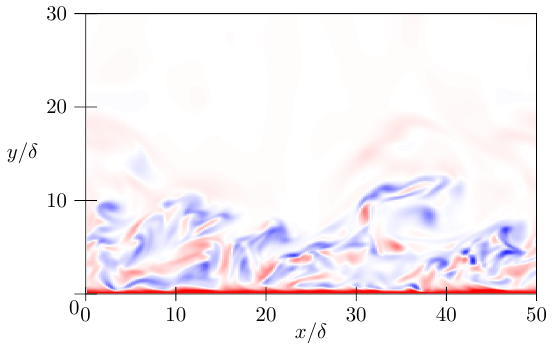} 
    \caption{$\omega t=0^{\circ}$}	
    \end{subfigure}\hfill
    \begin{subfigure}{0.49\linewidth}\centering 
        \includegraphics[width=2.5in]{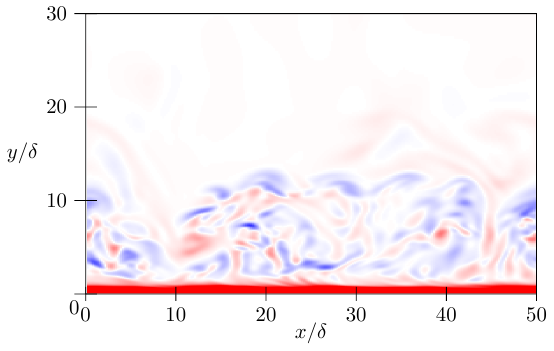}
    \caption{$\omega t=30^{\circ}$}	
    \end{subfigure}\hfill
    \begin{subfigure}{0.49\linewidth}\centering
        \includegraphics[width=2.5in]{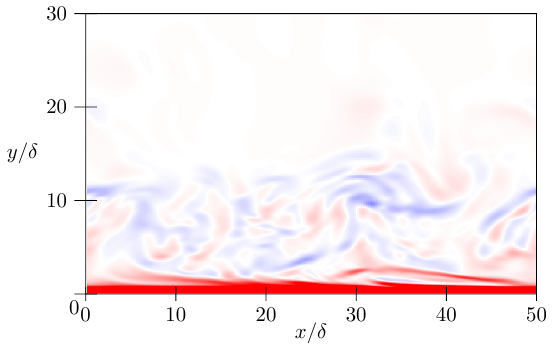}
    \caption{$\omega t=60^{\circ}$}	
    \end{subfigure}\hfill
    \begin{subfigure}{0.49\linewidth}\centering
        \includegraphics[width=2.5in]{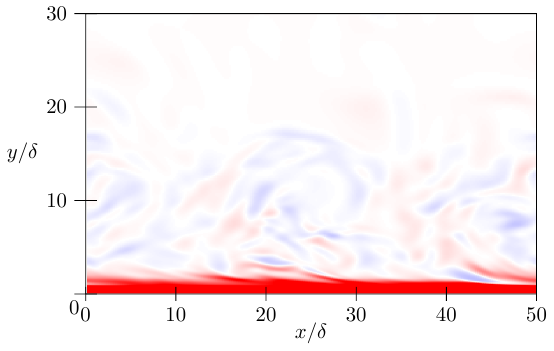}
    \caption{$\omega t=90^{\circ}$}	
    \end{subfigure}\hfill
    \begin{subfigure}{0.49\linewidth}\centering
        \includegraphics[width=2.5in]{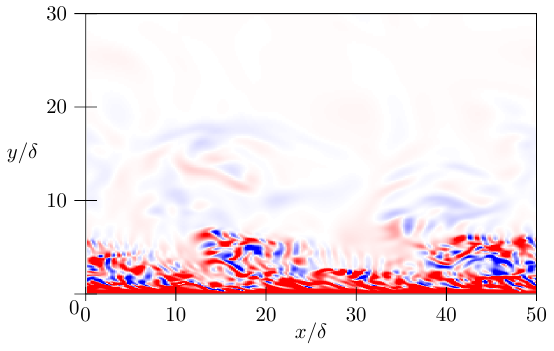}
    \caption{$\omega t=120^{\circ}$}	
    \end{subfigure}\hfill
    \begin{subfigure}{0.49\linewidth}\centering
        \includegraphics[width=2.5in]{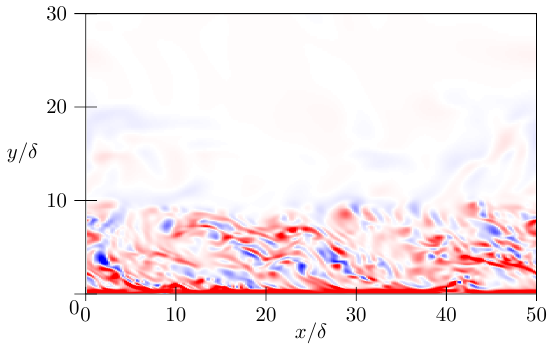}
    \caption{$\omega t=150^{\circ}$}	
    \end{subfigure}\hfill
    \caption{Normalized spanwise vorticity field in DNS of an OBL over an impermeable smooth wall at $\Rey_{\delta}=800$, for a smooth, impermeable wall. The eruption of velocity fluctuations during the decelerating portion of the cycle ($120^{\circ}$ and $150^{\circ}$) indicates that this flow is in the intermittent turbulent regime.}
    \label{fig:SP_OBL_Vorticity_800}
\end{figure}

Unlike the lower Reynolds number cases, figure \ref{fig:SP_OBL_Vorticity_800} shows significant vorticity throughout the cycle for the case at $\Rey_{\delta}=800$. Of particular interest is the range of scales seen at phase $120^{\circ}$, which is in the decelerating portion of the cycle. This eruption of velocity fluctuations, followed by partial relaminarization,  is characteristic of the intermittent turbulence regime. Similar observations were made by \citet{jensenTurbulentOscillatoryBoundary1989}, \citet{vittoriDirectSimulationTransition1998}, and \citet{salonNumericalInvestigationStokes2007}.

\begin{figure}
     \centering
     \begin{subfigure}{0.49\linewidth}\centering
          \includegraphics[width=2.75in]{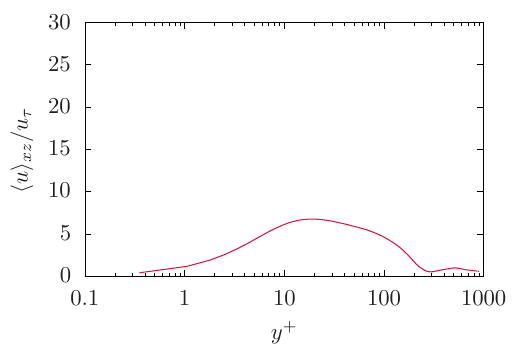}
          \caption{$\omega t = 0^{\circ}$}
     \end{subfigure}
     \begin{subfigure}{0.49\linewidth}\centering
          \includegraphics[width=2.75in]{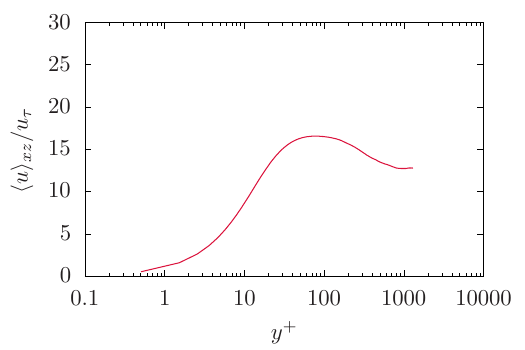}
          \caption{$\omega t = 30^{\circ}$}
     \end{subfigure}
     \begin{subfigure}{0.49\linewidth}\centering
          \includegraphics[width=2.75in]{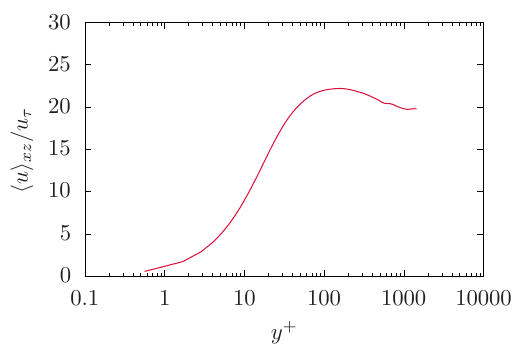}
          \caption{$\omega t = 60^{\circ}$}
     \end{subfigure}
     \begin{subfigure}{0.49\linewidth}\centering
          \includegraphics[width=2.75in]{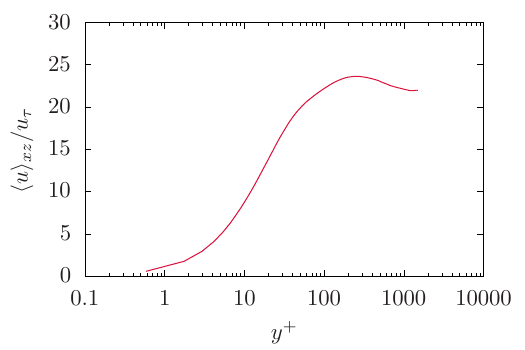}
          \caption{$\omega t = 90^{\circ}$}
     \end{subfigure}
     \begin{subfigure}{0.49\linewidth}\centering
          \includegraphics[width=2.75in]{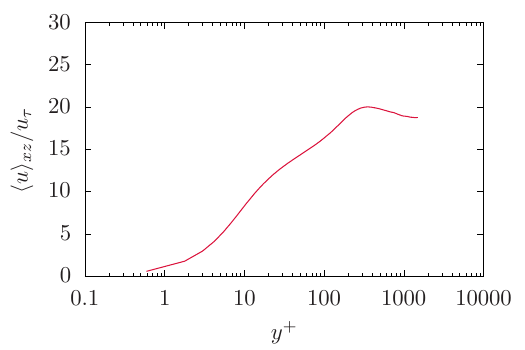}
          \caption{$\omega t = 120^{\circ}$}
     \end{subfigure}
     \begin{subfigure}{0.49\linewidth}\centering
          \includegraphics[width=2.75in]{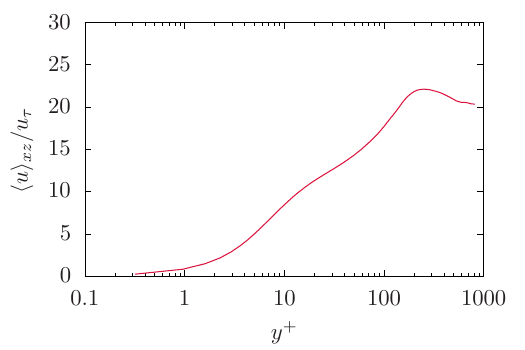}
          \caption{$\omega t = 150^{\circ}$}
     \end{subfigure}
     \caption{Wall scaled mean velocity profiles for $Re_{\delta} = 800$, for a smooth, impermeable wall. No logarithmic layer is observed.}
     \label{fig:SP Re 800 Log layer}
\end{figure}

{\color{revision}
While the vorticity field for \textcolor{revision}{$\Rey_\delta = 800$} shows a range of eddies, strict wall-bounded turbulence requires the existence of a logarithmic layer. To this end, we report in figure \ref{fig:SP Re 800 Log layer} vertical profiles of the spatially averaged streamwise velocity normalized using wall units, i.e., the friction velocity $u_{\tau} = \sqrt{\tau_{w}/\rho_{f}}$ as velocity scale and $\nu/u_{\tau}$ as length scale. Since none of the phases show a strict logarithmic layer, we conclude that the flow at $\Rey_\delta = 800$ is not fully turbulent, despite presenting significant fluctuations.
}
\section{\textcolor{revision}{Analysis of filter width}}
\label{sec:appendix_a}
{\color{revision}

In Eulerian-Lagrange simulations, the choice of filter width $\delta_f$ is an important modeling consideration. To justify the point-particle approximation $\delta_f$ must be much larger than the particle size $d_p$. Yet, resolving the vortical structures requires $\delta_f$ to also be as small as possible. For the present study, we chose $\delta_f=5d_p$. As recently shown in \citet{hausmannStudyDerivationClosures2024}, this filter size reduces errors associated with the point-particle approximation, while providing adequate resolution. 

\begin{figure}
\centering
 	\begin{subfigure}{0.49\linewidth}\centering
    \includegraphics[width=2.75in]{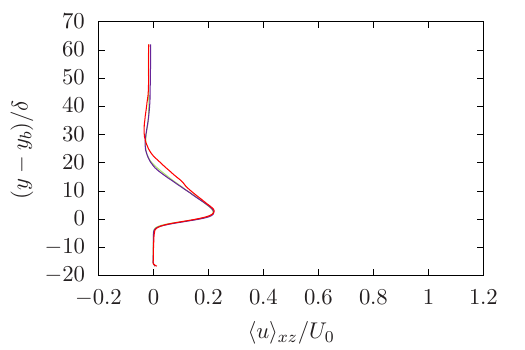}
    \caption{$\omega t = 0^\circ$}
    \end{subfigure}
   	\begin{subfigure}{0.49\linewidth}\centering
    \includegraphics[width=2.75in]{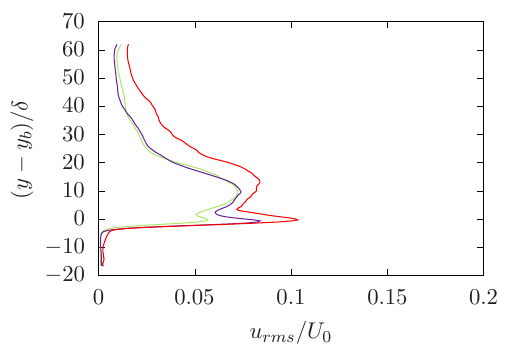}
    \caption{$\omega t = 0^\circ$}
    \end{subfigure}
    \begin{subfigure}{0.49\linewidth}\centering
    \includegraphics[width=2.75in]{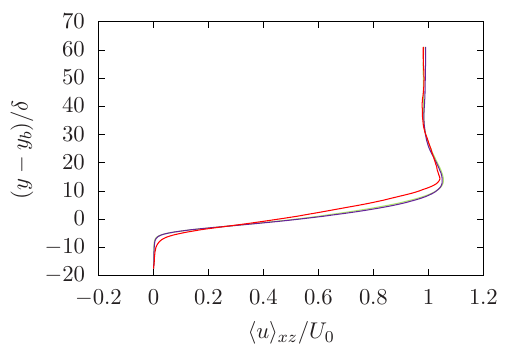}
    \caption{$\omega t = 90^\circ$}
    \end{subfigure}
    \begin{subfigure}{0.49\linewidth}\centering
    \includegraphics[width=2.75in]{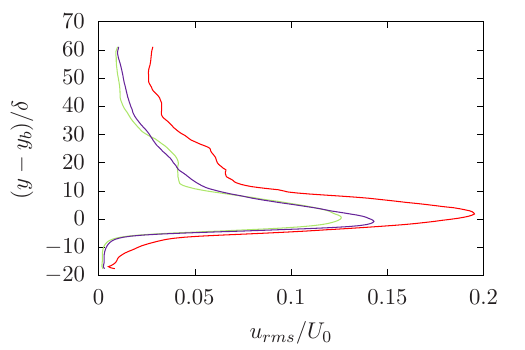}
    \caption{$\omega t = 90^\circ$}
    \end{subfigure}
    \begin{subfigure}{0.49\linewidth}\centering
    \includegraphics[width=2.75in]{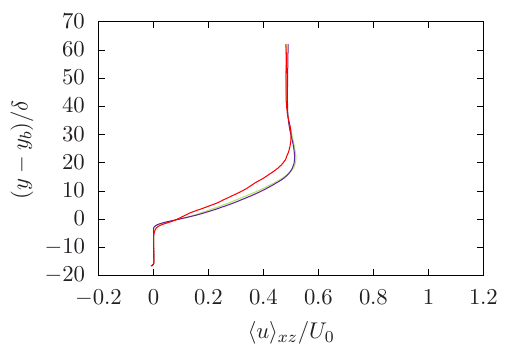}
    \caption{$\omega t = 150^\circ$}
    \end{subfigure}
    \begin{subfigure}{0.49\linewidth}\centering
    \includegraphics[width=2.75in]{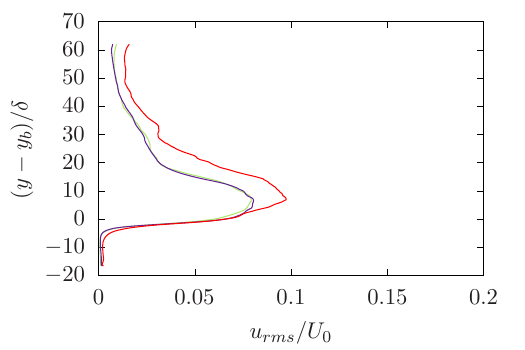}
    \caption{$\omega t = 150^\circ$}
    \end{subfigure}

 	\caption{\color{revision}  	Statistics of the streamwise fluid velocity for filter widths $\delta_f = 3d_p$(\textcolor{gpltphase150}{\LineStyleSolid}), \textcolor{revision}{$5d_p$}(\textcolor{gpltphase90}{\LineStyleSolid}), \textcolor{revision}{and $7d_p$}(\textcolor{gpltphase0}{\LineStyleSolid}): (a,c,e) mean and (b,d,f) rms fluctuations. There is little difference between results with $\delta_f=5d_p$ and $\delta_f=7d_p$, which indicates that either values are a good choice in Euler-Lagrange simulations.}
 	\label{fig:Filter_Width}
 \end{figure}
 
To show the impact of filter size on the fluid statistics, we performed additional simulations of the case at $\Rey_\delta = 800$ with filter sizes $\delta_f = 3d_p$, $5d_p$, and $7d_p$. Note that $\delta_f = 3d_p$ is generally considered to be too small to justify the point-particle approximation, while $\delta_f = 5d_p$ and $\delta_f = 7d_p$ are values commonly used in the literature \citet{capecelatroEulerLagrangeStrategy2013,hausmannStudyDerivationClosures2024}.

Figure \ref{fig:Filter_Width} shows the impact of varying the filter size on mean and rms fluctuations of the streamwise velocity at phases $0^\circ$, $90^\circ$, and $150^\circ$. We note that, as expected, $\delta_f = 3d_p$ shows the largest deviation. The results at $\delta_f = 5d_p$ and $\delta_f = 7d_p$ are sensibly similar, which justifies our choice of $\delta_f = 5d_p$ for the remaining simulations.

The reader interested in further details on the role of filter size in Euler-Langrange modeling is referred to \citet{hausmannStudyDerivationClosures2024}.
}

\section{\textcolor{revision2}{Review of the soft sphere collision model}}
\label{sec:appendix_c}
{\color{revision2}
The soft sphere collision model is dependent upon the parameters $k$ and $\eta$, which are the spring stiffness and dampening factor, respectively. They are related to the reduced mass $m_{ab} = (1/m_a + 1/m_b)^{-1}$, collision time $\tau_{\mathrm{col}}$, and coefficient of restitution $e$, as
\begin{eqnarray}
	k &=& \frac{m_{ab}}{\tau_{\mathrm{col}}^2}(\pi^2+\mathrm{ln}(e)^2), \\
	\eta &=& -2 \mathrm{ln}(e) \frac{\sqrt{m_{ab}k}}{\sqrt{\pi^2+\mathrm{ln}(e)^2}}.
\end{eqnarray}

The radius of influence, $\lambda$, allows us to robustly handle high-speed collisions by initiating the collision of high-speed particle pairs just before contact. Following \citet{finnParticleBasedModelling2016}, $\lambda$ is calculated as
\begin{equation}
	\lambda = \frac{\lambda_0}{2}(d_{p,a}+d_{p,b})\left ( \frac{\mathrm{CFL}^c_{ab}}{\mathrm{CFL}^c_{\mathrm{max}}} \right ),
\end{equation}
where the collisional CFL number is $\mathrm{CFL}^c_{ab} = (2 |\boldsymbol{u}_{ab,n}| \Delta t)/(d_{p}^a+d_{p}^b)$ and $\lambda_0$ is the maximum radius of influence permitted when the collision occurs at the maximum collision CFL number, $\mathrm{CFL}^c_\mathrm{max}$. The tangential collision force is modeled according to a static friction model,
\begin{equation}
	\boldsymbol{f}_{p,t}^\mathrm{c,b\rightarrow a} = -\mu_s |\boldsymbol{f}_{p,t}^\mathrm{c,b\rightarrow a} | \boldsymbol{t}_{ab}
\end{equation}
where $\boldsymbol{t}_{ab}$ is the tangential direction, and $\mu_s$ is the static friction coefficient. Collisions with walls are treated in the same way as above, but with the wall having infinite mass.
}

\section{\textcolor{revision}{Grid convergence study}}
\label{sec:appendix_d}
{\color{revision3}
In order to demonstrate grid independence, we conduct a simulation at $\Rey_\delta = 800$ with grid resolution $N_x \times N_y \times N_z = 672 \times 422 \times 134$, and compare the results to the simulation with resolution $N_x \times N_y \times N_z = 672 \times 211 \times 134$, for one  period. Thus the wall normal resolution is twice that in the coarser grid. Note that we do not refine the grid in the streamwise and spanwise directions because the gradients are primarily in wall normal direction, and the resolution in the streamwise and spanwise direction already exceeds the requirements for an oscillatory boundary layer over a smooth wall. That is, the resolution in the smooth wall case described in \ref{sec:appendix_SP} is $0.78 \delta$ and $0.39 \delta$ in the streamwise and spanwise directions, respectively, while it is $0.37 \delta$ in all directions for the bed cases. 

In figure \ref{fig:Grid_Con} we compare the fluid streamwise velocity profile, streamwise velocity fluctuations, and streamwise particle velocity.  Increasing the resolution does not lead to significant change in these statistics. This indicates that our resolution choice in this study is adequate up to $\Rey_\delta=800$.

\begin{figure}
    \centering
    \begin{subfigure}{0.49\linewidth}\centering
    \includegraphics[width=2.75in]{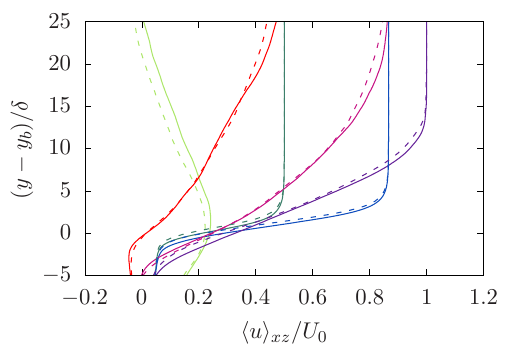}
    \caption{}
    \end{subfigure}
    \begin{subfigure}{0.49\linewidth}\centering
    \includegraphics[width=2.75in]{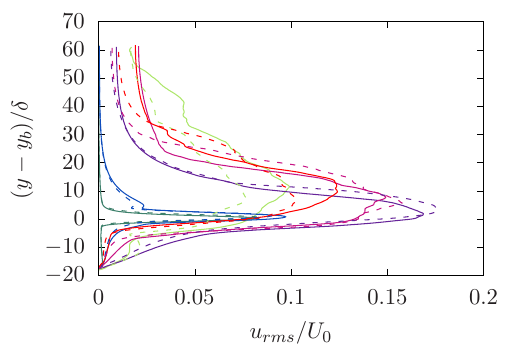}
    \caption{}
    \end{subfigure}
    \begin{subfigure}{0.49\linewidth}\centering
    \includegraphics[width=2.75in]{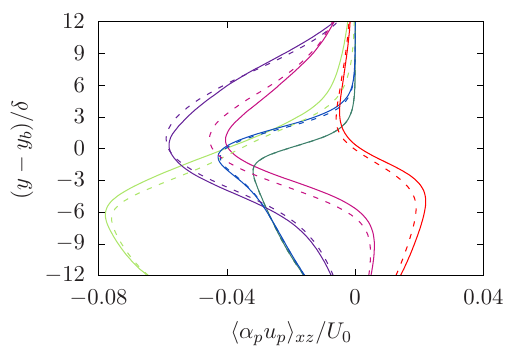}
    \caption{}
    \end{subfigure}
    \caption{\color{revision3} Profiles of mean fluid velocity and rms velocity fluctuations at $\Rey_\delta = 800$ Solid lines correspond to the grid described in the main document, while dashed lines correspond to the refined grid. Lines colors correspond to phases $\omega t =0$ (\textcolor{gpltphase0}{\LineStyleSolid}), $\omega t =30$ (\textcolor{gpltphase30}{\LineStyleSolid}), $\omega t =60$ (\textcolor{gpltphase60}{\LineStyleSolid}), $\omega t =90$ (\textcolor{gpltphase90}{\LineStyleSolid}), $\omega t =120$ (\textcolor{gpltphase120}{\LineStyleSolid}), and $\omega t =150$ (\textcolor{gpltphase150}{\LineStyleSolid}).. The agreement is close, showing that the simulation is grid converged.}
    \label{fig:Grid_Con}
\end{figure}

}

\section{Validation of the present Euler-Lagrange method for sediment transport}
\label{sec:appendix_E}
{\color{revision3}
To demonstrate the capability of the present model to capture accurate sediment transport, we reproduce numerically the experiments of \citet{aussillousInvestigationMobileGranular2013} of bedload transport in a channel flow.

\begin{figure}	\centering
  \begin{subfigure}{0.42\linewidth}
    \includegraphics[width=\linewidth]{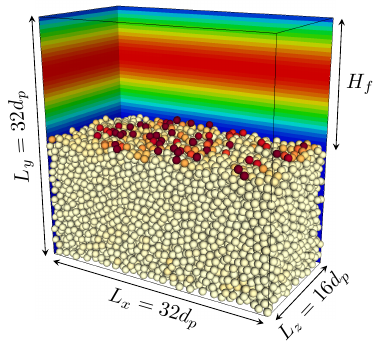}
    \caption{}
    \label{fig:sediment_validation_a}
  \end{subfigure}\hfill
  \begin{subfigure}{0.57\linewidth}
    \includegraphics[width=\linewidth]{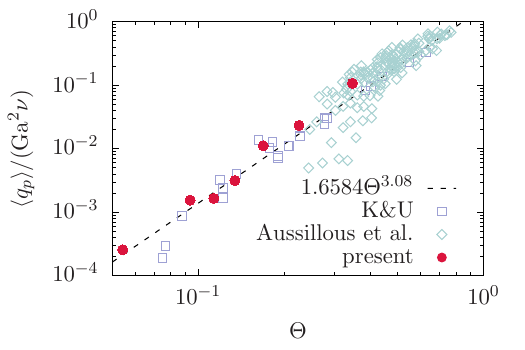}
    \caption{}
    \label{fig:sediment_validation_b}
  \end{subfigure}
  \caption{\color{revision3} Sediment transport in a laminar channel flow: (a) schematic of the configuration in our simulations and (b) variation of the normalized sediment flow rate with Shields number. The PR-DNS data is from \citep{kidanemariamInterfaceresolvedDirectNumerical2014}. The experimental data is from \citep{aussillousInvestigationMobileGranular2013}. Despite not capturing flow features at the particle scale, our method captures the variation of sediment transport with Shields number very well. }
  \label{fig:sediment_validation}
\end{figure}

Following earlier numerical effort in \citep{kidanemariamInterfaceresolvedDirectNumerical2014,charruSedimentTransportBedforms2016,raoCoarsegrainedModelingSheared2019}, we setup our numerical analogue as shown in figure \ref{fig:sediment_validation_a}. The sediment grains are monodisperse with diameter $d_p$, restitution coefficient $e=0.3$, and friction coefficient $\mu_c=0.4$ \citep{kidanemariamInterfaceresolvedDirectNumerical2014}. The domain dimensions are  $32d_p$ in the streamwise direction, $32d_p$ in the vertical direction, and $16d_p$ in the spanwise direction. We use a uniform grid with spacing $\Delta x=d_p/2$, same as in the main study.

The main non-dimensional numbers that control this problem are the bulk Reynolds number $\Rey=q_f/\nu$, where $q_f$ is the fluid flow rate, Galileo number $\mathrm{Ga}$, density ratio $\rho_p/\rho_f$, and Shields number $\Theta$. Following \citet{kidanemariamInterfaceresolvedDirectNumerical2014,charruSedimentTransportBedforms2016,raoCoarsegrainedModelingSheared2019} and to enable comparisons with \citet{aussillousInvestigationMobileGranular2013}, we define the Shields number based on the wall shear stress from a laminar Poiseuille flow. As discussed by \citet{kidanemariamInterfaceresolvedDirectNumerical2014}, this results in the following expression for the Shields number
\begin{equation}
  \Theta=\frac{6\Rey}{\mathrm{Ga}^2}\left(\frac{d_p}{H_f}\right)^2
\end{equation}
Note that we determine the fluid column height $H_f$ using the same approach outlined in \citep{kidanemariamInterfaceresolvedDirectNumerical2014} and \citep{raoCoarsegrainedModelingSheared2019}. 

Following \citet{kidanemariamInterfaceresolvedDirectNumerical2014}, we fix the bulk Reynolds number at $\Rey=375$, density ratio at $\rho_p/\rho_f=2.5$, and Galileo number at $\mathrm{Ga}=8.56$. We vary the Shields number from $0.05$ to $0.35$ by varying the bed height $H_b$. Similar to the main study, we drive the flow using a pressure gradient forcing, which is constant here. The flow reaches a steady state after about $200 H_b/U_b$, where $U_b$ is the bulk velocity, at which point we start collecting data.

Despite not resolving flow features at the particle scale, the methodology and computational methods outlined in this manuscript are well capable of reproducing meso-scale dynamics, including bedload sediment transport. This is well evidenced in figure \ref{fig:sediment_validation_b} which shows the variation of the normalized sediment $q_s/(\mathrm{Ga}^2\nu)$ flux with Shields number. The figure shows our data alongside data from the experiments of \citet{aussillousInvestigationMobileGranular2013}, PR-DNS of \citet{kidanemariamInterfaceresolvedDirectNumerical2014}, and their correlation $q_s/(\mathrm{Ga}^2\nu) = 1.6584 \Theta^{3.08}$. As clearly shown in figure \ref{fig:sediment_validation_b}, the present methodology yields good agreement with experimental and PR-DNS data. The variation of the normalized sediment flow rate is well captured across 3 orders of magnitude. It follows a cubic law, similar to the one determined by \citet{kidanemariamInterfaceresolvedDirectNumerical2014} and noted in other studies \citep{leightonViscousResuspension1986,charruInstabilityBedParticles2002,charruSedimentTransportBedforms2016}. This validates our approach for the problem in this study.

} 

\bibliography{References,References_houssem}

\begin{thebibliography}{64}
\expandafter\ifx\csname natexlab\endcsname\relax\def\natexlab#1{#1}\fi
\def\au#1{#1} \def\ed#1{#1} \def\yr#1{#1}\def\at#1{#1}\def\jt#1{\textit{#1}}
  \def\bt#1{#1}\def\bvol#1{\textbf{#1}} \def\vol#1{#1} \def\pg#1{#1}
  \def\publ#1{#1}\def\arxiv#1{#1}\def\org#1{#1}\def\st#1{\textit{#1}}

\bibitem[Akhavan {\em et~al.\/}(1991{\natexlab{{\em a\/}}})Akhavan, Kamm \&
  Shapiro]{akhavanInvestigationTransitionTurbulence1991}
{\sc \au{Akhavan, R.}, \au{Kamm, R.~D.} \& \au{Shapiro, A.~H.}}
  \yr{1991{\natexlab{{\em a\/}}}}  \at{An investigation of transition to
  turbulence in bounded oscillatory {{Stokes}} flows {{Part}} 1.
  {{Experiments}}}.  \jt{Journal of Fluid Mechanics}  \bvol{225},
  \pg{395--422}.

\bibitem[Akhavan {\em et~al.\/}(1991{\natexlab{{\em b\/}}})Akhavan, Kamm \&
  Shapiro]{akhavanInvestigationTransitionTurbulence1991a}
{\sc \au{Akhavan, R.}, \au{Kamm, R.~D.} \& \au{Shapiro, A.~H.}}
  \yr{1991{\natexlab{{\em b\/}}}}  \at{An investigation of transition to
  turbulence in bounded oscillatory {{Stokes}} flows {{Part}} 2. {{Numerical}}
  simulations}.  \jt{Journal of Fluid Mechanics}  \bvol{225},  \pg{423--444}.

\bibitem[Anderson \& Jackson(1967)]{andersonFluidMechanicalDescription1967}
{\sc \au{Anderson, T.~B.} \& \au{Jackson, R.}} \yr{1967}  \at{Fluid
  {{Mechanical Description}} of {{Fluidized Beds}}. {{Equations}} of
  {{Motion}}}.  \jt{Industrial \& Engineering Chemistry Fundamentals}
  \bvol{6}~(4),  \pg{527--539}.

\bibitem[Apte {\em et~al.\/}(2009)Apte, Martin \&
  Patankar]{apteNumericalMethodFully2009}
{\sc \au{Apte, Sourabh~V.}, \au{Martin, Mathieu} \& \au{Patankar, Neelesh~A.}}
  \yr{2009}  \at{A numerical method for fully resolved simulation ({{FRS}}) of
  rigid particle--flow interactions in complex flows}.  \jt{Journal of
  Computational Physics}  \bvol{228}~(8),  \pg{2712--2738}.

\bibitem[Arolla \& Desjardins(2015)]{arollaTransportModelingSedimenting2015}
{\sc \au{Arolla, S.~K.} \& \au{Desjardins, O.}} \yr{2015}  \at{Transport
  modeling of sedimenting particles in a turbulent pipe flow using
  {{Euler}}--{{Lagrange}} large eddy simulation}.  \jt{International Journal of
  Multiphase Flow}  \bvol{75},  \pg{1--11}.

\bibitem[Aussillous {\em et~al.\/}(2013)Aussillous, Chauchat, Pailha,
  M{\'e}dale \& Guazzelli]{aussillousInvestigationMobileGranular2013}
{\sc \au{Aussillous, Pascale}, \au{Chauchat, Julien}, \au{Pailha, Mickael},
  \au{M{\'e}dale, Marc} \& \au{Guazzelli, {\'E}lisabeth}} \yr{2013}
  \at{Investigation of the mobile granular layer in bedload transport by
  laminar shearing flows}.  \jt{Journal of Fluid Mechanics}  \bvol{736},
  \pg{594--615}.

\bibitem[Blondeaux \& Seminara(1979)]{blondeauxTransizioneIncipienteFondo1979}
{\sc \au{Blondeaux, P.} \& \au{Seminara, G.}} \yr{1979}  \at{{Transizione
  incipiente al fondo di un'onda di gravit{\`a}}}.  \jt{Atti della Accademia
  Nazionale dei Lincei. Classe di Scienze Fisiche, Matematiche e Naturali.
  Rendiconti}  \bvol{67}~(6),  \pg{408--417}.

\bibitem[Breugem(2012)]{breugemSecondorderAccurateImmersed2012}
{\sc \au{Breugem, Wim-Paul}} \yr{2012}  \at{A second-order accurate immersed
  boundary method for fully resolved simulations of particle-laden flows}.
  \jt{Journal of Computational Physics}  \bvol{231}~(13),  \pg{4469--4498}.

\bibitem[Breugem {\em et~al.\/}(2006)Breugem, Boersma \&
  Uittenbogaard]{breugemInfluenceWallPermeability2006}
{\sc \au{Breugem, W.~P.}, \au{Boersma, B.~J.} \& \au{Uittenbogaard, R.~E.}}
  \yr{2006}  \at{The influence of wall permeability on turbulent channel flow}.
   \jt{Journal of Fluid Mechanics}  \bvol{562},  \pg{35}.

\bibitem[Capecelatro \& Desjardins(2013{\natexlab{{\em
  a\/}}})]{capecelatroEulerLagrangeStrategy2013}
{\sc \au{Capecelatro, J.} \& \au{Desjardins, O.}} \yr{2013{\natexlab{{\em
  a\/}}}}  \at{An {{Euler}}--{{Lagrange}} strategy for simulating
  particle-laden flows}.  \jt{Journal of Computational Physics}  \bvol{238},
  \pg{1--31}.

\bibitem[Capecelatro \& Desjardins(2013{\natexlab{{\em
  b\/}}})]{capecelatroEulerianLagrangianModeling2013}
{\sc \au{Capecelatro, J.} \& \au{Desjardins, O.}} \yr{2013{\natexlab{{\em
  b\/}}}}  \at{Eulerian--{{Lagrangian}} modeling of turbulent liquid--solid
  slurries in horizontal pipes}.  \jt{International Journal of Multiphase Flow}
   \bvol{55},  \pg{64--79}.

\bibitem[Capecelatro {\em et~al.\/}(2014)Capecelatro, Pepiot \&
  Desjardins]{capecelatroNumericalCharacterizationModeling2014}
{\sc \au{Capecelatro, Jesse}, \au{Pepiot, Perrine} \& \au{Desjardins, Olivier}}
  \yr{2014}  \at{Numerical characterization and modeling of particle clustering
  in wall-bounded vertical risers}.  \jt{Chemical Engineering Journal}
  \bvol{245},  \pg{295--310}.

\bibitem[Carstensen {\em et~al.\/}(2010)Carstensen, Sumer \&
  Freds{\o}e]{carstensenCoherentStructuresWave2010}
{\sc \au{Carstensen, S.}, \au{Sumer, B.~M.} \& \au{Freds{\o}e, J.}} \yr{2010}
  \at{Coherent structures in wave boundary layers. {{Part}} 1. {{Oscillatory}}
  motion}.  \jt{Journal of Fluid Mechanics}  \bvol{646},  \pg{169--206}.

\bibitem[Carstensen {\em et~al.\/}(2012)Carstensen, Sumer \&
  Freds{\o}e]{carstensenNoteTurbulentSpots2012}
{\sc \au{Carstensen, S.}, \au{Sumer, B.~M.} \& \au{Freds{\o}e, J.}} \yr{2012}
  \at{A note on turbulent spots over a rough bed in wave boundary layers}.
  \jt{Physics of Fluids}  \bvol{24}~(11),  \pg{115104}.

\bibitem[Charru {\em et~al.\/}(2016)Charru, Bouteloup, Bonometti \&
  Lacaze]{charruSedimentTransportBedforms2016}
{\sc \au{Charru, F.}, \au{Bouteloup, J.}, \au{Bonometti, T.} \& \au{Lacaze,
  L.}} \yr{2016}  \at{Sediment transport and bedforms: A numerical study of
  two-phase viscous shear flow}.  \jt{Meccanica}  \bvol{51}~(12),
  \pg{3055--3065}.

\bibitem[Charru \&
  {Mouilleron-Arnould}(2002)]{charruInstabilityBedParticles2002}
{\sc \au{Charru, Fran{\c c}ois} \& \au{{Mouilleron-Arnould}, H{\'e}l{\`e}ne}}
  \yr{2002}  \at{Instability of a bed of particles sheared by a viscous flow}.
  \jt{Journal of Fluid Mechanics}  \bvol{452},  \pg{303--323}.

\bibitem[Conley \& Inman(1994)]{conleyVentilatedOscillatoryBoundary1994}
{\sc \au{Conley, Daniel~C.} \& \au{Inman, Douglas~L.}} \yr{1994}
  \at{Ventilated oscillatory boundary layers}.  \jt{Journal of Fluid Mechanics}
   \bvol{273},  \pg{261--284}.

\bibitem[Costamagna {\em et~al.\/}(2003)Costamagna, Vittori \&
  Blondeaux]{costamagnaCoherentStructuresOscillatory2003}
{\sc \au{Costamagna, P.}, \au{Vittori, G.} \& \au{Blondeaux, P.}} \yr{2003}
  \at{Coherent structures in oscillatory boundary layers}.  \jt{Journal of
  Fluid Mechanics}  \bvol{474},  \pg{1--33}.

\bibitem[Dave \& Kasbaoui(2023)]{daveMechanismsDragReduction2023}
{\sc \au{Dave, Himanshu} \& \au{Kasbaoui, M.~Houssem}} \yr{2023}
  \at{Mechanisms of drag reduction by semidilute inertial particles in
  turbulent channel flow}.  \jt{Physical Review Fluids}  \bvol{8}~(8),
  \pg{084305}.

\bibitem[Finn \& Li(2016)]{finnRegimesSedimentturbulenceInteraction2016}
{\sc \au{Finn, J.~R.} \& \au{Li, M.}} \yr{2016}  \at{Regimes of
  sediment-turbulence interaction and guidelines for simulating the multiphase
  bottom boundary layer}.  \jt{International Journal of Multiphase Flow}
  \bvol{85},  \pg{278--283}.

\bibitem[Finn {\em et~al.\/}(2016)Finn, Li \&
  Apte]{finnParticleBasedModelling2016}
{\sc \au{Finn, J.~R.}, \au{Li, M.} \& \au{Apte, S.~V.}} \yr{2016}  \at{Particle
  based modelling and simulation of natural sand dynamics in the wave bottom
  boundary layer}.  \jt{Journal of Fluid Mechanics}  \bvol{796},
  \pg{340--385}.

\bibitem[Fytanidis {\em et~al.\/}(2021)Fytanidis, Garc{\'i}a \&
  Fischer]{fytanidisMeanFlowStructure2021}
{\sc \au{Fytanidis, D.~K.}, \au{Garc{\'i}a, M.~H.} \& \au{Fischer, P.~F.}}
  \yr{2021}  \at{Mean flow structure and velocity--bed shear stress maxima
  phase difference in smooth wall, transitionally turbulent oscillatory
  boundary layers: Direct numerical simulations}.  \jt{Journal of Fluid
  Mechanics}  \bvol{928},  \pg{A33}.

\bibitem[Ghodke \& Apte(2016)]{ghodkeDNSStudyParticlebedturbulence2016}
{\sc \au{Ghodke, C.~D.} \& \au{Apte, S.~V.}} \yr{2016}  \at{{{DNS}} study of
  particle-bed-turbulence interactions in an oscillatory wall-bounded flow}.
  \jt{Journal of Fluid Mechanics}  \bvol{792},  \pg{232--251}.

\bibitem[Ghodke \& Apte(2018)]{ghodkeRoughnessEffectsSecondorder2018}
{\sc \au{Ghodke, C.~D.} \& \au{Apte, S.~V.}} \yr{2018}  \at{Roughness effects
  on the second-order turbulence statistics in oscillatory flows}.
  \jt{Computers \& Fluids}  \bvol{162},  \pg{160--170}.

\bibitem[Gibilaro {\em et~al.\/}(2007)Gibilaro, Gallucci, Di~Felice \&
  Pagliai]{gibilaroApparentViscosityFluidized2007}
{\sc \au{Gibilaro, L.~G.}, \au{Gallucci, K.}, \au{Di~Felice, R.} \&
  \au{Pagliai, P.}} \yr{2007}  \at{On the apparent viscosity of a fluidized
  bed}.  \jt{Chemical Engineering Science}  \bvol{62}~(1),  \pg{294--300}.

\bibitem[Hausmann {\em et~al.\/}(2024)Hausmann, Ch{\'e}ron, Evrard \& van
  Wachem]{hausmannStudyDerivationClosures2024}
{\sc \au{Hausmann, Max}, \au{Ch{\'e}ron, Victor}, \au{Evrard, Fabien} \&
  \au{van Wachem, Berend}} \yr{2024}  \at{Study and derivation of closures in
  the volume-filtered framework for particle-laden flows}.  \jt{Journal of
  Fluid Mechanics}  \bvol{996},  \pg{A41}.

\bibitem[Hino {\em et~al.\/}(1976)Hino, Sawamoto \&
  Takasu]{hinoExperimentsTransitionTurbulence1976}
{\sc \au{Hino, M.}, \au{Sawamoto, M.} \& \au{Takasu, S.}} \yr{1976}
  \at{Experiments on transition to turbulence in an oscillatory pipe flow}.
  \jt{Journal of Fluid Mechanics}  \bvol{75}~(2),  \pg{193--207}.

\bibitem[Hsu {\em et~al.\/}(2004)Hsu, Jenkins \&
  Liu]{hsuTwoPhaseSedimentTransport2004}
{\sc \au{Hsu, Tian-Jian}, \au{Jenkins, James~T.} \& \au{Liu, Philip L.~F.}}
  \yr{2004}  \at{On {{Two-Phase Sediment Transport}}: {{Sheet Flow}} of
  {{Massive Particles}}}.  \jt{Proceedings: Mathematical, Physical and
  Engineering Sciences}  \bvol{460}~(2048),  \pg{2223--2250},  \arxiv{arXiv:
  4143214}.

\bibitem[Jensen {\em et~al.\/}(1989)Jensen, Sumer \&
  Freds{\o}e]{jensenTurbulentOscillatoryBoundary1989}
{\sc \au{Jensen, B.~L.}, \au{Sumer, B.~M.} \& \au{Freds{\o}e, J.}} \yr{1989}
  \at{Turbulent oscillatory boundary layers at high {{Reynolds}} numbers}.
  \jt{Journal of Fluid Mechanics}  \bvol{206},  \pg{265--297}.

\bibitem[Jewel {\em et~al.\/}(2019)Jewel, Fujisawa \&
  Murakami]{jewelEffectSeepageFlow2019}
{\sc \au{Jewel, A.}, \au{Fujisawa, K.} \& \au{Murakami, A.}} \yr{2019}
  \at{Effect of seepage flow on incipient motion of sand particles in a bed
  subjected to surface flow}.  \jt{Journal of Hydrology}  \bvol{579},
  \pg{124178}.

\bibitem[Kasbaoui(2019)]{kasbaouiTurbulenceModulationSettling2019}
{\sc \au{Kasbaoui, M.~H.}} \yr{2019}  \at{Turbulence modulation by settling
  inertial aerosols in {{Eulerian-Eulerian}} and {{Eulerian-Lagrangian}}
  simulations of homogeneously sheared turbulence}.  \jt{Physical Review
  Fluids}  \bvol{4}~(12),  \pg{124308}.

\bibitem[Kasbaoui \&
  Herrmann(2025)]{kasbaouiHighfidelityMethodologyParticleresolved2025}
{\sc \au{Kasbaoui, M.~Houssem} \& \au{Herrmann, Marcus}} \yr{2025}  \at{A
  high-fidelity methodology for particle-resolved direct numerical
  simulations}.  \jt{International Journal of Multiphase Flow}  \bvol{187},
  \pg{105175}.

\bibitem[Kasbaoui {\em et~al.\/}(2019)Kasbaoui, Koch \&
  Desjardins]{kasbaouiClusteringEulerEuler2019}
{\sc \au{Kasbaoui, M.~Houssem}, \au{Koch, Donald~L.} \& \au{Desjardins,
  Olivier}} \yr{2019}  \at{Clustering in {{Euler}}--{{Euler}} and
  {{Euler}}--{{Lagrange}} simulations of unbounded homogeneous particle-laden
  shear}.  \jt{Journal of Fluid Mechanics}  \bvol{859},  \pg{174--203}.

\bibitem[Kempe \& Fr{\"o}hlich(2012)]{kempeImprovedImmersedBoundary2012}
{\sc \au{Kempe, Tobias} \& \au{Fr{\"o}hlich, Jochen}} \yr{2012}  \at{An
  improved immersed boundary method with direct forcing for the simulation of
  particle laden flows}.  \jt{Journal of Computational Physics}
  \bvol{231}~(9),  \pg{3663--3684}.

\bibitem[Kidanemariam \&
  Uhlmann(2014)]{kidanemariamInterfaceresolvedDirectNumerical2014}
{\sc \au{Kidanemariam, A.~G.} \& \au{Uhlmann, M.}} \yr{2014}
  \at{Interface-resolved direct numerical simulation of the erosion of a
  sediment bed sheared by laminar channel flow}.  \jt{International Journal of
  Multiphase Flow}  \bvol{67},  \pg{174--188}.

\bibitem[Leighton \& Acrivos(1986)]{leightonViscousResuspension1986}
{\sc \au{Leighton, David} \& \au{Acrivos, Andreas}} \yr{1986}  \at{Viscous
  resuspension}.  \jt{Chemical Engineering Science}  \bvol{41}~(6),
  \pg{1377--1384}.

\bibitem[Liu {\em et~al.\/}(1996)Liu, Davis \&
  Downing]{liuWaveinducedBoundaryLayer1996}
{\sc \au{Liu, Philip L.-F.}, \au{Davis, Matthew~H.} \& \au{Downing, Sean}}
  \yr{1996}  \at{Wave-induced boundary layer flows above and in a permeable
  bed}.  \jt{Journal of Fluid Mechanics}  \bvol{325},  \pg{195--218}.

\bibitem[Maxey \& Riley(1983)]{maxeyEquationMotionSmall1983a}
{\sc \au{Maxey, Martin~R.} \& \au{Riley, James~J.}} \yr{1983}  \at{Equation of
  motion for a small rigid sphere in a nonuniform flow}.  \jt{The Physics of
  Fluids}  \bvol{26}~(4),  \pg{883--889}.

\bibitem[Mazzuoli {\em et~al.\/}(2020)Mazzuoli, Blondeaux, Vittori, Uhlmann,
  Simeonov \& Calantoni]{mazzuoliInterfaceresolvedDirectNumerical2020}
{\sc \au{Mazzuoli, M.}, \au{Blondeaux, P.}, \au{Vittori, G.}, \au{Uhlmann, M.},
  \au{Simeonov, J.} \& \au{Calantoni, J.}} \yr{2020}  \at{Interface-resolved
  direct numerical simulations of sediment transport in a turbulent oscillatory
  boundary layer}.  \jt{Journal of Fluid Mechanics}  \bvol{885}.

\bibitem[Mazzuoli {\em et~al.\/}(2016)Mazzuoli, Kidanemariam, Blondeaux,
  Vittori \& Uhlmann]{mazzuoliFormationSedimentChains2016}
{\sc \au{Mazzuoli, M.}, \au{Kidanemariam, A.~G.}, \au{Blondeaux, P.},
  \au{Vittori, G.} \& \au{Uhlmann, M.}} \yr{2016}  \at{On the formation of
  sediment chains in an oscillatory boundary layer}.  \jt{Journal of Fluid
  Mechanics}  \bvol{789},  \pg{461--480}.

\bibitem[Mazzuoli {\em et~al.\/}(2019)Mazzuoli, Kidanemariam \&
  Uhlmann]{mazzuoliDirectNumericalSimulations2019}
{\sc \au{Mazzuoli, Marco}, \au{Kidanemariam, Aman~G.} \& \au{Uhlmann, Markus}}
  \yr{2019}  \at{Direct numerical simulations of ripples in an oscillatory
  flow}.  \jt{Journal of Fluid Mechanics}  \bvol{863},  \pg{572--600}.

\bibitem[Mazzuoli \& Vittori(2019)]{mazzuoliTurbulentSpotsOscillatory2019}
{\sc \au{Mazzuoli, M.} \& \au{Vittori, G.}} \yr{2019}  \at{Turbulent spots in
  an oscillatory flow over a rough wall}.  \jt{European Journal of Mechanics -
  B/Fluids}  \bvol{78},  \pg{161--168}.

\bibitem[{Meza-Valle} \& Pujara(2022)]{meza-valleFlowOscillatoryBoundary2022}
{\sc \au{{Meza-Valle}, Claudio} \& \au{Pujara, Nimish}} \yr{2022}  \at{Flow in
  oscillatory boundary layers over permeable beds}.  \jt{Physics of Fluids}
  \bvol{34}~(9),  \pg{092112}.

\bibitem[O'Donoghue \& Wright(2004)]{odonoghueFlowTunnelMeasurements2004}
{\sc \au{O'Donoghue, Tom} \& \au{Wright, Scott}} \yr{2004}  \at{Flow tunnel
  measurements of velocities and sand flux in oscillatory sheet flow for
  well-sorted and graded sands}.  \jt{Coastal Engineering}  \bvol{51}~(11),
  \pg{1163--1184}.

\bibitem[Ozdemir {\em et~al.\/}(2014)Ozdemir, Hsu \&
  Balachandar]{ozdemirDirectNumericalSimulations2014}
{\sc \au{Ozdemir, C.}, \au{Hsu, T.~J.} \& \au{Balachandar, S.}} \yr{2014}
  \at{Direct numerical simulations of transition and turbulence in
  smooth-walled {{Stokes}} boundary layer}.  \jt{Physics of Fluids}  \bvol{26}.

\bibitem[Pedocchi {\em et~al.\/}(2011)Pedocchi, Cantero \&
  Garc{\'i}a]{pedocchiTurbulentKineticEnergy2011}
{\sc \au{Pedocchi, F.}, \au{Cantero, M.~I.} \& \au{Garc{\'i}a, M.~H.}}
  \yr{2011}  \at{Turbulent kinetic energy balance of an oscillatory boundary
  layer in the transition to the fully turbulent regime}.  \jt{Journal of
  Turbulence}  \bvol{12},  \pg{N32}.

\bibitem[Rao \& Capecelatro(2019)]{raoCoarsegrainedModelingSheared2019}
{\sc \au{Rao, Arun~Ashok} \& \au{Capecelatro, Jesse}} \yr{2019}
  \at{Coarse-grained modeling of sheared granular beds}.  \jt{International
  Journal of Multiphase Flow}  \bvol{114},  \pg{258--267}.

\bibitem[Saffman(1965)]{saffmanLiftSmallSphere1965}
{\sc \au{Saffman, P.~G.}} \yr{1965}  \at{The lift on a small sphere in a slow
  shear flow}.  \jt{Journal of Fluid Mechanics}  \bvol{22}~(2),  \pg{385--400}.

\bibitem[Salon {\em et~al.\/}(2007)Salon, Armenio \&
  Crise]{salonNumericalInvestigationStokes2007}
{\sc \au{Salon, S.}, \au{Armenio, V.} \& \au{Crise, A.}} \yr{2007}  \at{A
  numerical investigation of the {{Stokes}} boundary layer in the turbulent
  regime}.  \jt{Journal of Fluid Mechanics}  \bvol{570},  \pg{253--296}.

\bibitem[Sarpkaya(1993)]{sarpkayaCoherentStructuresOscillatory1993}
{\sc \au{Sarpkaya, T.}} \yr{1993}  \at{Coherent structures in oscillatory
  boundary layers}.  \jt{Journal of Fluid Mechanics}  \bvol{253},
  \pg{105--140}.

\bibitem[Scott \& Kilgour(1969)]{scottDensityRandomClose1969}
{\sc \au{Scott, G.~D.} \& \au{Kilgour, D.~M.}} \yr{1969}  \at{The density of
  random close packing of spheres}.  \jt{Journal of Physics D: Applied Physics}
   \bvol{2}~(6),  \pg{863--866}.

\bibitem[Shuai {\em et~al.\/}(2022)Shuai, Dhas, Roy \&
  Kasbaoui]{shuaiInstabilityDustyVortex2022}
{\sc \au{Shuai, Shuai}, \au{Dhas, Darish~Jeswin}, \au{Roy, Anubhab} \&
  \au{Kasbaoui, M.~Houssem}} \yr{2022}  \at{Instability of a dusty vortex}.
  \jt{Journal of Fluid Mechanics}  \bvol{948},  \pg{A56}.

\bibitem[Shuai \& Kasbaoui(2022)]{shuaiAcceleratedDecayLamb2022}
{\sc \au{Shuai, Shuai} \& \au{Kasbaoui, M.~Houssem}} \yr{2022}  \at{Accelerated
  decay of a {{Lamb}}--{{Oseen}} vortex tube laden with inertial particles in
  {{Eulerian}}--{{Lagrangian}} simulations}.  \jt{Journal of Fluid Mechanics}
  \bvol{936}.

\bibitem[Shuai {\em et~al.\/}(2024)Shuai, Roy \&
  Kasbaoui]{shuaiMergerCorotatingVortices2024}
{\sc \au{Shuai, Shuai}, \au{Roy, Anubhab} \& \au{Kasbaoui, M.~Houssem}}
  \yr{2024}  \at{The merger of co-rotating vortices in dusty flows}.
  \jt{Journal of Fluid Mechanics}  \bvol{981},  \pg{A27}.

\bibitem[Sparrow {\em et~al.\/}(2012)Sparrow, Pokrajac \& van~der
  A]{sparrowEFFECTBEDPERMEABILITY2012}
{\sc \au{Sparrow, Kathryn}, \au{Pokrajac, Dubravka} \& \au{van~der A,
  Dominic~A.}} \yr{2012}  \at{{{THE EFFECT OF BED PERMEABILITY ON OSCILLATORY
  BOUNDARY LAYER FLOW}}}.  \jt{Coastal Engineering Proceedings} ~(33),
  \pg{26--26}.

\bibitem[Stokes(1855)]{stokesEffectsInternalFriction1855}
{\sc \au{Stokes, G.~G.}} \yr{1855}  \at{On the effects of internal friction of
  fluids on the motion of pendulums}.  \jt{Transactions of the Cambridge
  Philosophical Society}  \bvol{9}.

\bibitem[Tenneti {\em et~al.\/}(2011)Tenneti, Garg \&
  Subramaniam]{tennetiDragLawMonodisperse2011}
{\sc \au{Tenneti, S.}, \au{Garg, R.} \& \au{Subramaniam, S.}} \yr{2011}
  \at{Drag law for monodisperse gas--solid systems using particle-resolved
  direct numerical simulation of flow past fixed assemblies of spheres}.
  \jt{International Journal of Multiphase Flow}  \bvol{37}~(9),
  \pg{1072--1092}.

\bibitem[Uhlmann(2005)]{uhlmannImmersedBoundaryMethod2005}
{\sc \au{Uhlmann, Markus}} \yr{2005}  \at{An immersed boundary method with
  direct forcing for the simulation of particulate flows}.  \jt{Journal of
  Computational Physics}  \bvol{209}~(2),  \pg{448--476}.

\bibitem[Van~Doren \& Kasbaoui(2024)]{vandorenTurbulenceModulationDense2024}
{\sc \au{Van~Doren, Jonathan~S.} \& \au{Kasbaoui, M.~Houssem}} \yr{2024}
  \at{Turbulence modulation in dense liquid-solid channel flow}.  \jt{Physical
  Review Fluids}  \bvol{9}~(6),  \pg{064306}.

\bibitem[Vittori {\em et~al.\/}(2020)Vittori, Blondeaux, Mazzuoli, Simeonov \&
  Calantoni]{vittoriSedimentTransportOscillatory2020}
{\sc \au{Vittori, G.}, \au{Blondeaux, P.}, \au{Mazzuoli, M.}, \au{Simeonov, J.}
  \& \au{Calantoni, J.}} \yr{2020}  \at{Sediment transport under oscillatory
  flows}.  \jt{International Journal of Multiphase Flow}  \bvol{133},
  \pg{103454}.

\bibitem[Vittori \& Verzicco(1998)]{vittoriDirectSimulationTransition1998}
{\sc \au{Vittori, G.} \& \au{Verzicco, R.}} \yr{1998}  \at{Direct simulation of
  transition in an oscillatory boundary layer}.  \jt{Journal of Fluid
  Mechanics}  \bvol{371},  \pg{207--232}.

\bibitem[Voermans {\em et~al.\/}(2017)Voermans, Ghisalberti \&
  Ivey]{voermansVariationFlowTurbulence2017}
{\sc \au{Voermans, J.~J.}, \au{Ghisalberti, M.} \& \au{Ivey, G.~N.}} \yr{2017}
  \at{The variation of flow and turbulence across the sediment--water
  interface}.  \jt{Journal of Fluid Mechanics}  \bvol{824},  \pg{413--437}.

\bibitem[Xiong {\em et~al.\/}(2020)Xiong, Qi, Gao, Xu, Ren \&
  Cheng]{xiongBypassTransitionMechanism2020}
{\sc \au{Xiong, Chengwang}, \au{Qi, Xiang}, \au{Gao, Ankang}, \au{Xu, Hui},
  \au{Ren, Chengjiao} \& \au{Cheng, Liang}} \yr{2020}  \at{The bypass
  transition mechanism of the {{Stokes}} boundary layer in the intermittently
  turbulent regime}.  \jt{Journal of Fluid Mechanics}  \bvol{896},  \pg{A4}.

\bibitem[Yuan \& Madsen(2014)]{yuanExperimentalStudyTurbulent2014}
{\sc \au{Yuan, Jing} \& \au{Madsen, {\relax Ole.S}.}} \yr{2014}
  \at{Experimental study of turbulent oscillatory boundary layers in an
  oscillating water tunnel}.  \jt{Coastal Engineering}  \bvol{89},
  \pg{63--84}.

\end{thebibliography}
\end{document}